\documentclass[epj]{svjour}
\usepackage{graphics}

\usepackage[english]{babel}
\usepackage[dvipdfm]{graphicx}
\usepackage[T1]{fontenc}
\usepackage[latin1]{inputenc}
\usepackage{verbatim}
\usepackage{amsfonts}
\usepackage{hyperref}
\usepackage{color}
\usepackage{epsfig}
\usepackage{setspace}
\usepackage{amsmath}
\usepackage{amstext}
\usepackage{amsbsy}
\usepackage{amscd}
\usepackage{amsgen}
\usepackage{amsopn}
\usepackage{amsxtra}
\usepackage{amssymb}
\usepackage{mathrsfs}
\usepackage{euscript}
\usepackage{url}
\usepackage[bottom]{footmisc}
\usepackage{footnote}
\usepackage[numbers, sort&compress]{natbib} % per i riferimenti bibliografici
\usepackage[font=footnotesize,hang]{caption} % per le note

\usepackage{numprint} % per la notazione scientifica dei numeri
\usepackage{eso-pic}
\usepackage{subfigure} % per inserire sottofigure
\usepackage{captcont}
\usepackage{multirow} % per inserire tabelle con multi righe
\usepackage{epigraph} % per inserire epigrafi
\usepackage{latexsym}
\usepackage{units}
\usepackage{xspace}
\usepackage{vmargin}
\usepackage{colortbl}
\usepackage{rotating}

\usepackage[switch*]{lineno}
\linenumbers
\newcommand{\lto}[1]{\longrightarrow#1}

\newcommand{\sign}{\text{sign}}
\renewcommand{\(}{\left(}
\renewcommand{\)}{\right)}
\renewcommand{\[}{\left[}
\renewcommand{\]}{\right]}

\date{}
\DeclareGraphicsExtensions{.eps}

\usepackage{fancyhdr}

\begin{document}
\graphicspath{{figure/}}

\title{HERMES: Simulating the Propagation of Ultra-High Energy Cosmic Rays\thanks{This paper is based on the author's PhD thesis, that was awarded the INFN Bruno Rossi Prize in 2012.}}

\author{Manlio De Domenico\inst{1}\inst{2}\inst{3}% etc
% \thanks is optional - remove next line if not needed
\thanks{\emph{Present address:} Departament d'Enginyeria Inform\'atica i Matem\'atiques, Universidad Rovira i Virgili, Avda. Paisos Catalans 26, 43007 Tarragona, Spain}%
}                     % Do not remove
%
%\offprints{}          % Insert a name or remove this line
%

\institute{Laboratorio sui Sistemi Complessi, Scuola Superiore di Catania, Via Valdisavoia 9, 95123 Catania, Italy \and Dipartimento di Fisica e Astronomia, Universit\'a degli Studi di Catania, Via S. Sofia 64, 95123 Catania, Italy \and Istituto Nazionale di Fisica Nucleare, Sez. di Catania, Via S. Sofia 64, 95123 Catania, Italy}
\date{Received: date / Revised version: date}
% The correct dates will be entered by Springer
%
\abstract{
The study of ultra-high energy cosmic rays (UHECR) at Earth cannot prescind from the study of their propagation in the Universe. In this paper, we present HERMES, the \emph{ad hoc} Monte Carlo code we have developed for the realistic simulation of UHECR propagation. We discuss the modeling adopted to simulate the cosmology, the magnetic fields, the interactions with relic photons and the production of secondary particles. In order to show the potential applications of HERMES for astroparticle studies, we provide an estimation of the surviving probability of UHE protons, the GZK horizons of nuclei and the all-particle spectrum observed at Earth in different astrophysical scenarios. Finally, we show the expected arrival direction distribution of UHECR produced from nearby candidate sources.
A stable version of HERMES will be released in the next future for public use together with libraries of already propagated nuclei to allow the community to perform mass composition and energy spectrum analysis with our simulator.
 % end of PACS codes
} %end of abstract

\maketitle

\section{Introduction}

A final answer about the origin and the composition of ultra-high energy cosmic rays (UEHCR) is still missing. Several models have been proposed for the acceleration of UHECR \cite{hillas1984origin, hill1987ultra, berezinsky1997cosmic, berezinsky1997ultrahigh, venkatesan1997constraints, farrar1998correlation, fargion1999ultra, arons2003magnetars} (see \cite{nagano2000observations, bhattacharjee2000origin} and Ref. therein for a review) and it is generally accepted that the candidate sources are extragalactic and trace the distribution of luminous matter on large scales \cite{waxman1997signature}. The recent result reported by the Pierre Auger Collaboration, from observations in the southern hemisphere, experimentally supports compact sources with a number density in the range $10^{-5}$-$10^{-3}$~Mpc$^{-3}$\,\cite{dedomenico2011bounds,auger2013bounds}, showing a correlation between the observed data with energy above 57~EeV and the distribution of nearby mass distribution \cite{auger2010correlation}. Observations in the northern hemisphere by the HiRes Collaboration, but with smaller statistics and a different energy scale, do not confirm this result \cite{abbasi2008search}, while more recent measurements by the Telescope Array Collaboration, based on 25 observed events with energy larger than 57~EeV, suggest a correlation with nearby Active Galactic Nuclei with chance probability of 2\% \cite{abu2012search}.

Even the observed suppression of UHECR, due to their propagation in the Universe, is still debated: in fact, UHECR of extragalactic origin with energy above 100\,EeV (1\,EeV $= 10^{18}$ eV) could be subjected to a strong attenuation because of their interaction with relic photons of the extragalactic background radiation. Recently, the Pierre Auger Collaboration reported a suppression of the spectrum above 40\,EeV with significance greater than 20 standard deviations \cite{settimo2012spectrum,abraham2010measurement}, improving previous measurements \cite{abraham2008observation,abbasi2008first}. Such results are compatible with the existence of the GZK effect \cite{greisen1966end,zatsepin1966upper}, although not providing a definitive evidence. In fact, alternative suitable scenarios, compatible with the same observations, involve a spectrum cutoff directly at the source.

%\cite{abraham2008observation} and the HiRes Collaboration \cite{abbasi2008first} reported the experimental evidence of the suppression of the UHECRs spectrum with a statistical significance of about six and five standard deviations, respectively. More recent observations reveal a suppression of the spectrum above 40 EeV with significance greater than 20 standard deviations \cite{abraham2010measurement}. Such results are compatible with the existence of the GZK effect \cite{greisen1966end,zatsepin1966upper}, although not providing a definitive evidence. In fact, alternative suitable scenarios, compatible with the same observations, involve a spectrum cutoff directly at the source.

It is evident that both modeling and realistic simulations of production and propagation mechanisms are required to shed light on the nature of UHECR \cite{puget1976photonuclear,allard2005uhe,harari2006ultrahigh,hooper2007intergalactic,allard2008implications,globus2008propagation,allard2009interactions}, trying to avoid the limitations \cite{kachelriess2009gzk} given by the continuous energy loss approximation adopted by some authors to simplify calculations.

In this study, we present the general structure of our propagation code (HERMES) \cite{dedomenico2011thesis}. We show the simulated diffusion of charged particles in both turbulent and structured magnetic fields for energy values ranging from $10^{17}$~eV to $10^{21}$~eV and we provide an estimation of mean free paths and energy-loss lengths of UHE nuclei. The expected GZK horizon is reported together with a comparison with existing results and an estimation of the expected spectrum at Earth is compared against recent observations.

%%%%%%%%%%%%%%%%%%%%%%%%%%%%%%%%%%%%%%%%%%%
%%%%%%%%%%%%%%%%%%%%%%%%%%%%%%%%%%%%%%%%%%%

\section{Simulating the propagation of UHECR with HERMES: background radiation and magnetic fields}

In this section, we describe the HERMES propagation code, presenting the modeling adopted for i) the cosmological framework, ii) the cosmic background radiation (microwave, infrared/optical and radio), iii) the regular component of the Galactic magnetic field and the irregular component of both the Galactic and the extragalactic magnetic fields, iv) the cross sections describing the interactions between UHE nuclei and photons of extragalactic background radiation, v) the production of secondary particles because of such interactions. In the following, we will briefly describe such a framework, to provide the reader with the necessary tools to understand the parameterizations and the energy-loss equation adopted in our Monte Carlo code.

%%%%%%%%%%%%%%%%%%%%%%%%%%%%%%%%%%%%%%%%%%%
%%%%%%%%%%%%%%%%%%%%%%%%%%%%%%%%%%%%%%%%%%%
\subsection{Cosmological framework}\label{sec-cosmology}

Motivated by up-to-date observations, we have chosen a general Friedmann's Universe, defined by a Friedmann-Robertson-Walker metric, to be the cosmological framework in HERMES. Let us consider the Einstein equation in the classical General Relativity framework to describe the gravitational field. Under the assumptions of an isotropic and homogeneous Universe, we also consider the Friedmann-Robertson-Walker (FRW) metric
\begin{eqnarray}
ds^{2}=c^{2}dt^{2}-a^{2}(t)\[ \frac{dr^{2}}{1-\kappa r^{2}}+r^{2}\( d\theta^{2}+\sin^{2}\theta d\phi^{2} \) \],
\end{eqnarray}
where $a(t)$ is the scale factor, such that $a(0)=1$ is its present value, while the parameter $\kappa$ accounts for the spatial curvature: $\kappa=-1$ denotes an open metric, $\kappa=0$ a flat metric and $\kappa=1$ a closed metric. Indeed, we consider the Universe as a perfect fluid with energy density $\varrho$ and pressure $p$, described by the stress-energy tensor $T_{\mu\nu}=\(\varrho+\frac{p}{c^{2}}\)u_{\mu}u_{\nu}+pg_{\mu\nu}$, where $u_{\mu}$ denotes the 4-velocity. Friedmann equations can be derived from such assumptions. 

HERMES is able to propagate particles in a $\Lambda$CDM Universe, with several tunable parameters expressed in terms of the critical density $\varrho_{c}=3H^{2}/8\pi G$. More specifically, we consider $\Omega_{b}$ due to baryonic matter, $\Omega_{c}$ due to cold dark matter, $\Omega_{\Lambda}$ due to dark energy, $\Omega_{r}$ due to radiation and $\Omega_{\kappa}$ for the spatial curvature. If we define the redshift $z$ by $1+z=a^{-1}(t)$, the first Friedmann equation can be written in terms of $z$ and of density parameters as
\begin{eqnarray}
\label{def-Hz}
\frac{H^{2}(z)}{H_{0}^{2}}=\Omega_{r}(1+z)^{4}+\Omega_{M}(1+z)^{3}+\Omega_{k}(1+z)^{2}+\Omega_{\Lambda},
\end{eqnarray}
where $\Omega_{M}=\Omega_{b}+\Omega_{c}$ is the total density of matter and $H_{0}$ is the Hubble parameter at the present time. By taking into account that the radiation density contributes only in the early Universe, i.e. at high redshifts, whereas in practice it is negligible in the late Universe, the constraint $\Omega_{M}+\Omega_{\kappa}+\Omega_{\Lambda}=1$ for the density parameters can be obtained from very general considerations. We will describe further in text the role of Eq.\,(\ref{def-Hz}) in the numerical simulation of the propagation of UHECR.

It is worth noticing that a particle with energy $E(z)$ at redshift $z$, propagating through the Universe and not subjected to energy loss processes, will adiabatically lose its energy because of the expansion of the Universe (of course, by assuming a cosmological model where the Universe is expanding), and it will be observed with energy $E_{0}=E/(1+z)$ at the Earth.

The values of all relevant parameters discussed so far, as the Hubble constant, the density of matter and energy, can be freely varied in our simulator to reproduce very different cosmological models, and a study of the impact of cosmology on the GZK horizon of UHECR protons has been recently published \cite{dedomenico2012influence}.

%%%%%%%%%%%%%%%%%%%%%%%%%%%%%%%%%%%%%%%%%%%
%%%%%%%%%%%%%%%%%%%%%%%%%%%%%%%%%%%%%%%%%%%

\subsection{Spectrum of UHECR and Evolution of sources}

Let $Q(E)$ indicate the injection spectrum of UHECR at the source, representing the number of particles injected per unit energy and time, and let us indicate the source luminosity by
\begin{eqnarray}
\mathcal{L}=\int_{E_{\text{min}}}^{E_{\text{max}}}Q(E)EdE,
\end{eqnarray}
quantifying the energy emitted from the source in terms of UHECR per unit time. Here, we are assuming that UHECR at the source can be produced from a minimum energy $E_{\text{min}}$ to a maximum energy $E_{\text{max}}$. There are some arguments predicting a power-law injection spectrum of both Galactic and extragalactic CRs \cite{fermi1949origin,bell1974upper,torres2004astrophysical}. Under such an assumption, we can rewrite the injection spectrum as a function of the source luminosity by $Q(E)=\mathcal{L}\mathcal{N}E^{-\gamma}$, being $\gamma$ the injection index and $\mathcal{N}$ a normalization factor. The source luminosity may increase with redshift, as well as the comoving density of sources: in general such a cosmological source evolution depends on several factors, related to the class of astrophysical sources under consideration. If the source evolution is present, the luminosity should include an additional factor $\mathcal{H}(z)=(1+z)^{m}$, giving $\mathcal{L}(z)=\mathcal{H}(z)\mathcal{L}$. It is worth remarking that the source evolution factor can play a significant role for the study of the energy spectrum of UHECR at Earth. Thus, in general, the injection spectrum simulated in HERMES is given by $Q(z,E)=\mathcal{H}(z)Q(0,E)$. The following evolution factors are available in our simulator:
\begin{enumerate}
\item Star formation rate (SFR) \cite{hopkins2006normalization}:
\begin{eqnarray}
\mathcal{H}_{\text{SFR}}(z)=\left\{\begin{array}{ll}
(1+z)^{3.4} & z<1,\\
2^{3.7}(1+z)^{-0.3} & 1<z<4,\\
2^{3.7}\times5^{3.2}(1+z)^{-3.5} & z>4;
\end{array}\right.
\end{eqnarray}

\item Gamma-ray burst (GRB) \cite{yuksel2007enhanced}: $\mathcal{H}_{\text{GRB}}(z)=(1+z)^{1.4}\mathcal{H}_{\text{SFR}}(z)$; 

\item Active galactic nuclei (AGN) \cite{stanev2009ultra,hasinger2005luminosity}: 
\begin{eqnarray}
\mathcal{H}_{\text{AGN}}(z)=\left\{\begin{array}{ll}
(1+z)^{5} & z<1.7,\\
2.7^{5} & 1.7<z<2.7,\\
2.7^{5}\times10^{0.43(2.7-z)} & z>2.7;
\end{array}\right.
\end{eqnarray}

\item Quasi-stellar object (QSO) \cite{engel2001neutrinos}: 
\begin{eqnarray}
\mathcal{H}_{\text{QSO}}(z)=\left\{\begin{array}{ll}
(1+z)^{3} & z<1.9,\\
(1+1.9)^{3} & 1.9<z<2.7,\\
(1+1.9)^{3}e^{1-z/2.7} & z>2.7.
\end{array}\right.
\end{eqnarray}

\end{enumerate}

In the case of a uniform evolution $\mathcal{H}_{\text{unif}}(z)=(1+z)^{3}$, whereas in the case of no evolution $\mathcal{H}(z)=1$ can be assumed.

%%%%%%%%%%%%%%%%%%%%%%%%%%%%%%%%%%%%%%%%%%%
%%%%%%%%%%%%%%%%%%%%%%%%%%%%%%%%%%%%%%%%%%%

\subsection{Modeling the extragalactic background radiation}\label{sec-ebr-parameterization}

The propagation of UHECRs is affected by their interactions with photons of the extragalactic background radiation (EBR). While the relevant energy losses will be discussed successively in the text, we briefly describe here the models of background radiations simulated in HERMES.

In fact, EBR modeling is rather difficult, if the well known cosmic microwave background is excluded. Such a radiation should be produced by the assembly of matter into stars and galaxies, as well as by the evolution of such systems which releases radiant energy powered by gravitational and nuclear processes. Absorption of large frequency radiation by dust and re-emission at small frequency considerably increase the infrared component of the background light, whose investigation should shed light on structure formations processes. In the following, we indicate with $\epsilon$ the relic photon energy in eV, $n(\epsilon)$ the photon spectral number density in units of photons cm$^{-3}$~eV$^{-1}$ and $\epsilon^{2}n(\epsilon)$ the energy density in units of eV~cm$^{-3}$.

EBR spans over almost 20 decades, according to observations and models, from radio waves around $10^{-7}$~eV up to the high energy $\gamma-$ray photons of several GeV, with cosmic microwave background (CMB), the relic blackbody radiation from the Big Bang, being the dominant form of electromagnetic energy followed by ultraviolet/optical (CUVOB) and infrared backgrounds (CIRB). 

For the propagation of UHECR nuclei, in HERMES we adopt the blackbody model with temperature $T_{0}\simeq 2.725$~K for CMB. The semi-analytical ``model D''  proposed by Finke et al \cite{finke2010modeling}, modeling the star formation rate recently introduced by Hopkins and Beacom \cite{hopkins2006normalization} is adopted for CIOB (for a more detailed treatment of infrared and optical background radiations we refer to \cite{hauser2001cosmic,malkan2001empirically,lagache2003modelling,stecker2006intergalactic,stecker2008spectrum,franceschini2008extragalactic,finke2010modeling} and Refs. therein). The model proposed in \cite{protheroe1996new} is adopted for the universal radio background (URB).

\begin{figure}[!t]
\centering
\includegraphics[width=0.70\textwidth]{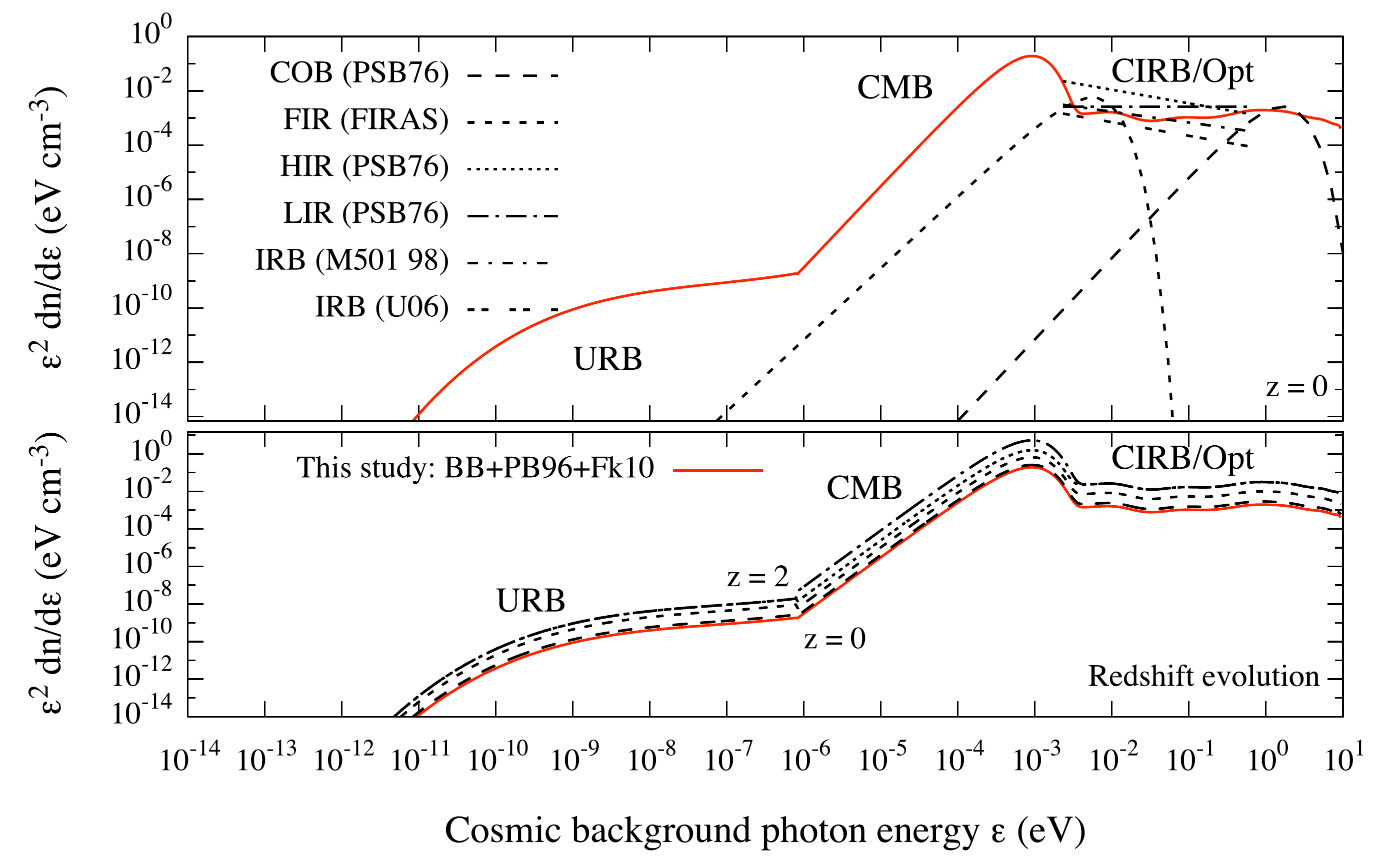}
    \caption{Top panel: different parameterizations of extragalactic background radiation as a function of relic photon energy: CMB, (Far, Low and High) IRB and COB. The red line indicates the EBL parameterization included in HERMES. The other parameterizations, shown for reference, are taken from PSB76 \cite{puget1976photonuclear}, FIRAS \cite{fixsen2011probing}, ER98 \cite{epele1998propagation}, Mkn501-98 \cite{funk1998upper}, U06 \cite{uryson2006ultra}. Lower panel: evolution of EBR for different values of redshift. Photon energy is considered in the laboratory frame.}
    \label{fig:ebl-model}
\end{figure}

Some models of extragalactic background radiations are shown in the top panel of Fig.\,\ref{fig:ebl-model}, as a function of the photon energy $\epsilon$ in the laboratory frame. The red solid line indicates the EBL parameterization included in HERMES, and it should be considered the default, where not specified otherwise. For sake of completeness, we also show the common parameterizations by Puget, Stecker and Bredekamp (PSB76) for COB, lower and higher IRB (LIR and HIR, respectively) \cite{puget1976photonuclear}, and other IRB models, derived from theoretical arguments or experimental observations \cite{epele1998propagation,funk1998upper,uryson2006ultra,fixsen2011probing}. The bottom panel of the same figure shows the evolution with redshift for different values of $z$, ranging from 0 to 2.

By assuming that the cosmological model of gravitation is described by general relativity and electromagnetism by Maxwell theory, a theoretical consequence of the adiabatic expansion of the Universe is that photons should propagate along null geodesics and that the CMB temperature should evolve with redshift as $T(z)=T_{0} (1+z)^{1-\beta}$, with $\beta=0$. From the same arguments, it can be shown that the energy of CMB photons evolve as $E(z)=E_{0} (1+z)$, whereas their number density evolve as $n(\epsilon,z)=n(\epsilon,z=0)(1+z)^{3}$.

The evolution of the density of CIRB photons is still debated and depends on the adopted scenario for the luminosity evolution. Two models, included in HERMES, have been recently suggested by Stecker et al \cite{stecker2006intergalactic}:
\begin{enumerate}
\item \textbf{Base-line model}: 
\begin{eqnarray}
\mathcal{E}(z)=\left\{
\begin{array}{ll}
(1+z)^{3.1} & z\leq 1.3\\
(1+1.3)^{3.1} & 1.3< z\leq 6\\
0 & z>6
\end{array}\right.
\end{eqnarray}
\item \textbf{Fast model}: 
\begin{eqnarray}
\mathcal{E}(z)=\left\{
\begin{array}{ll}
(1+z)^{4} & z\leq 1\\
(1+1)^{4} & 1.3< z\leq 6\\
0 & z>6
\end{array}\right.
\end{eqnarray}
\end{enumerate}
In the current cosmological epoch and at the IRB maximum epoch, which is around $z=2$, the fast evolution model provides an higher density than base-line model. In any case, it is worth remarking that the cosmological evolution of the infrared background density is much slower than that of CMB.

Finally, the evolution of the density of CRB photons included in HERMES is the one proposed by Protheroe and Biermann, who modified the luminosity evolution to fit the source counts \cite{protheroe1996new}:
\begin{eqnarray}
\mathcal{E}(z)=\left\{
\begin{array}{ll}
(1+z)^{4} & z< 0.8\\
(1+0.8)^{4} &  z\geq 0.8
\end{array}\right.
\end{eqnarray}
where the value $z_{0}=0.8$ has been obtained from the best fit for both normal galaxies and radio galaxies.

We will see further in this chapter that the radio background is negligible when the propagation of high energy nuclei is considered: conversely, it plays an important role during the propagation of high energy photons.

%%%%%%%%%%%%%%%%%%%%%%%%%%%%%%%%%%%%%%%%%%%
%%%%%%%%%%%%%%%%%%%%%%%%%%%%%%%%%%%%%%%%%%%

\subsection{Modeling magnetic fields}

The presence of magnetic fields, both in the intergalactic space and in our galaxy, has a non-negligible impact on the propagation of charged nuclei. It is thus of fundamental importance to investigate the structure of galactic and extragalactic magnetic field (GMF and EMF, respectively), that have a direct impact on the energy spectrum, the strength of the anisotropy signal and the correlation with candidate sources. In order to simulate the diffusion of particles with charge $q=Ze$ in magnetic fields, we adopt in HERMES a standard approach, based on the numerical integration of the equation of motions obtained in the ultra-relativistic approximation, in the case of nuclei.

If the electric field is absent (or negligible) and we assume the case of a particle in ultra-relativistic regime, i.e. the particle travels at the speed of light in the direction $\hat{v}(t)$ at time $t$ subjected to a magnetic field $\vec{B}(\vec{r})$ along the trajectory $\vec{r}(t)$, the Lorentz equation reduces to the set of six ordinary differential equations defined by
\begin{eqnarray}
\frac{d\vec{r}(t)}{dt} &=& c\hat{v}(t)\nonumber\\
\frac{d\hat{v}(t)}{dt} &=& \frac{qc^{2}}{E}\hat{v}(t)\wedge\vec{B}(\vec{r}).
\end{eqnarray}
In practice, charged particles accelerating in a magnetic field lose energy because of the emission of synchrotron radiation: in the case of light particles as electrons or positrons, such energy loss should be taken into account during the propagation, whereas for heavier particles as protons it is negligible. 

While the trajectory of a charged particle along the regular field is deterministic, i.e. for a given initial condition only one solution to the equations of motion exists, the trajectory of a particle through the turbulent field is stochastic, thus not unique, and it depends on the features of the irregular field as its r.m.s. strength and its coherence length. Unfortunately, we have no exact knowledge of both galactic and extragalactic magnetic fields and, as a consequence, the investigation of charged particles propagation through our galaxy and intergalactic space, respectively, should be based either on empirical or theoretical models and numerical simulations. For the simulation of the irregular component of the magnetic field, we adopt in HERMES the approach proposed by Giacalone and Jokipii \cite{giacalone1994charged,giacalone1999transport}, based on a local step-by-step simulation of the turbulent field. 

\subsubsection{Simulating a turbulent magnetic field} 

The randomness of the irregular component of the magnetic field is probably due to the evolution of stochastic fluctuations which are correlated up to a given correlation scale. In fact, such an irregular component should show the features typical of correlated flows undergoing turbulent evolution, characterized by a minimum and a maximum scale of turbulence, $\ell_{\text{min}}$ and $\ell_{\text{max}}$, respectively. The particles scatter off the magnetic irregularities and change their pitch angle $\theta$, but not their velocity. The pitch angle scattering is principally dominated by the inhomogeneities with scales of the order of the Larmor radius, i.e. by resonance, providing an effective mechanism of isotropization as long as $r_{L} < \ell_{\text{max}}$.

Our simulation of such an irregular behavior is based on the following approach. The turbulent magnetic field $\vec{B}(\vec{r})$ satisfies two main requirements: i) it is a zero-mean field $\langle \vec{B}(\vec{r}) \rangle=0$ with ii) non-vanishing fluctuations $\langle\vec{B}^{2}(\vec{r})\rangle=B_{\text{rms}}>0$. Let $\vec{k}$ be the wave vectors with modulus $k$, power spectrum $\mathcal{P}(k)\propto k^{-5/3}$ and amplitudes $\vec{B}(\vec{k})$ of its Fourier modes following the Kolmogorov spectrum $|\vec{B}(\vec{k})|^{2}\propto k^{-11/3}$: such a field defines a turbulent Kolmogorov 3D magnetic field\footnotemark\footnotetext{The spectral index is 8/3 and 5/3 for 2D and 1D magnetic fields, respectively.}. In the Fourier space the wave vectors satisfy $\frac{2\pi}{\ell_{\text{max}}}\leq k\leq \frac{2\pi}{\ell_{\text{min}}}$, where the correlation length of the field is equal to \cite{harari2002lensing}
\begin{eqnarray}
\Lambda_{c}=\frac{1}{2}\ell_{\text{max}}\frac{\gamma-1}{\gamma}\frac{1-(\ell_{\text{min}}/\ell_{\text{max}})^{\gamma}}{1-(\ell_{\text{min}}/\ell_{\text{max}})^{\gamma-1}},
\end{eqnarray}
where $\gamma$ is the spectral index of the Kolmogorov spectrum. The approach, proposed by Giacalone and Jokipii \cite{giacalone1994charged,giacalone1999transport} considers the field as the sum of $N_{m}$ modes, physically corresponding to the superposition of a finite number of plane waves:
\begin{eqnarray}
\vec{B}(\vec{r})=\sum_{n=1}^{N_{m}}A_{n}\hat{\varepsilon}_{n}e^{i\vec{k}_{n}\cdot\vec{r}+i\beta_{n}},
\end{eqnarray}
where $\hat{\varepsilon}_{n}=\cos\alpha_{n}\hat{x}_{n}+i\sin\alpha_{n}\hat{y}_{n}$ and the amplitude $A_{n}$ of the $n-$th plane wave is given by
\begin{eqnarray}
A^{2}_{n}=\mathcal{A}B^{2}_{\text{irr}}G(\vec{k}),
\end{eqnarray}
with 
\begin{eqnarray}
G(\vec{k})=\frac{\Delta V_{n}}{1+(k\Lambda_{c})^{\gamma}},\quad \Delta V_{n}=4\pi k^{2}\Delta k,\quad \mathcal{A}=\(\sum_{n=1}^{N_{m}}G(\vec{k}_{n})\)^{-1}.
\end{eqnarray}
In this last equation, the index $\gamma$ is equal to 11/3, 8/3 and 5/3 for 3D, 2D and 1D turbulent magnetic fields, respectively. The direction of the $n-$th wave vector $\hat{k}_{n}$ is randomly chosen: the unit vectors $\hat{x}_{n}$ and $\hat{y}_{n}$ are chosen in order to form an orthogonal basis with $\hat{k}_{n}$ and the real numbers $\alpha_{n}$ and $\beta_{n}$ represent random polarizations and phases, respectively. For practical applications, the spacing $\Delta k$ between $k_{\text{min}}=\frac{2\pi}{\ell_{\text{max}}}$ and $k_{\text{max}}=\frac{2\pi}{\ell_{\text{min}}}$ should be constant in logarithmic scale and the number of modes $N_{m}$ should be large enough to obtain the expected results in the small-angle regime. The main advantage of such an approach is the definition of the turbulent field at any point in space with arbitrary precision at the price of a much slower computation than other methods. Where not otherwise specified, in the following we will make use of the isotropic model, although the simulation of the composite model is also allowed by our code. Moreover, we will consider a total magnetic field $\vec{B}=\vec{B}_{\text{tot}}=\vec{B}_{\text{reg}}+\vec{B}_{\text{irr}}$ and, following Ref.\,\cite{casse2001transport}, we define the turbulence level by
\begin{eqnarray}
\eta=\frac{\langle\vec{B}^{2}\rangle}{B_{\text{reg}}^{2}+\langle\vec{B}^{2}\rangle}.
\end{eqnarray}

The above arguments can be used to simulate the turbulent component of both the extragalactic and the Galactic magnetic fields. For instance, the deflection $\delta_{\text{irr}}$ due to the irregular component of the Galactic magnetic field can be estimated by assuming that the particle undergoes a brownian motion at the scale of the coherence length $\Lambda$ of the field and that the ratio between the traversed distance $D$ and $\Lambda$ provides an estimation of the number of magnetic regions traversed \cite{roulet2004astroparticle,giacinti2011ultrahigh}:
\begin{eqnarray}
\delta_{\text{irr}}=\frac{1}{\sqrt{2}}\frac{ZeB_{\text{rms}}}{E}\(D\Lambda\)^{\frac{1}{2}}\simeq 0.6^{\circ}\frac{10^{20}~\text{eV}}{E/Z}\frac{B_{\text{rms}}}{4~\mu\text{G}}\(\frac{D}{3~\text{kpc}}\)^{\frac{1}{2}}\(\frac{\Lambda}{50~\text{pc}}\)^{\frac{1}{2}},
\end{eqnarray}
being $B_{\text{rms}}=\langle B^{2}_{\text{irr}}\rangle$. Similarly, in the case of the extragalactic magnetic field, by considering the appropriate coherence length and by neglecting energy loss processes \cite{bhattacharjee2000origin} we obtain
\begin{eqnarray}
\delta_{\text{irr}}=\simeq 0.8^{\circ}\frac{10^{20}~\text{eV}}{E/Z}\frac{B_{\text{rms}}}{1~\text{nG}}\(\frac{D}{10~\text{Mpc}}\)^{\frac{1}{2}}\(\frac{\Lambda}{1~\text{Mpc}}\)^{\frac{1}{2}}.
\end{eqnarray}

\subsubsection{Simulating the Galactic magnetic field} 

In spiral galaxies, the turbulent component of the magnetic field is almost always strongest within the spiral arms, following the distribution of cool gas and dust, whereas the regular component is generally weak within spiral arms, except for rare cases like M51 with strong density waves. However, the regular field also extends far into the inter-arm regions. Observations suggest that the large-scale spiral field produce an halo, extending outside the galactic disks. In cylindrical coordinates, the distribution of the magnetic field $B(\rho, \phi, z)$ in the galaxy can be described by the product of three separated components, related to pure radial dependence $R(\rho)$, spiral ``winding'' modulation $S(\rho, \phi)$, and halo extinction $H(z)$, respectively. Several models have been proposed to describe the regular component of the magnetic field in our galaxy. A detailed description of all models is beyond the scope of the present paper, therefore we limit to mention the models included in HERMES.
%In the following we briefly describe the most common ones, which are also the models included in HERMES.
The structure of the magnetic field obtained by dynamo mechanisms can be described by modes of different azimuthal symmetry in the disk, and vertical symmetry perpendicular to the disk plane: bisymmetric (BSS) or axisymmetric (ASS), depending on $\pi$ or $2\pi$ symmetry, respectively. Along the vertical dimension, the field can change direction while traversing the disk plane (odd or A-parity) or keep it fixed (even or S-parity). Thus, the possible patterns of the spiral field are four, indicated with the notation BSS-S, BSS-A, ASS-S and ASS-A, and they are all present in HERMES, coupled with galactic magnetic field models proposed by Stanev \cite{stanev1997ultra}, Harari, Mollerach and Roulet (HMR) \cite{harari1999toes}, and Tinyakov and Tkachev \cite{tinyakov2002tracing}. For the sake of completeness, we refer to Refs. \cite{prouza2003galactic,kachelriess2007galactic,page2007three,sun2008radio,jansson2009constraining} for other models, which we plan to include in HERMES, describing the galactic magnetic field.

\begin{figure}[!t]
  \begin{center} 
     \includegraphics[width=7.5cm]{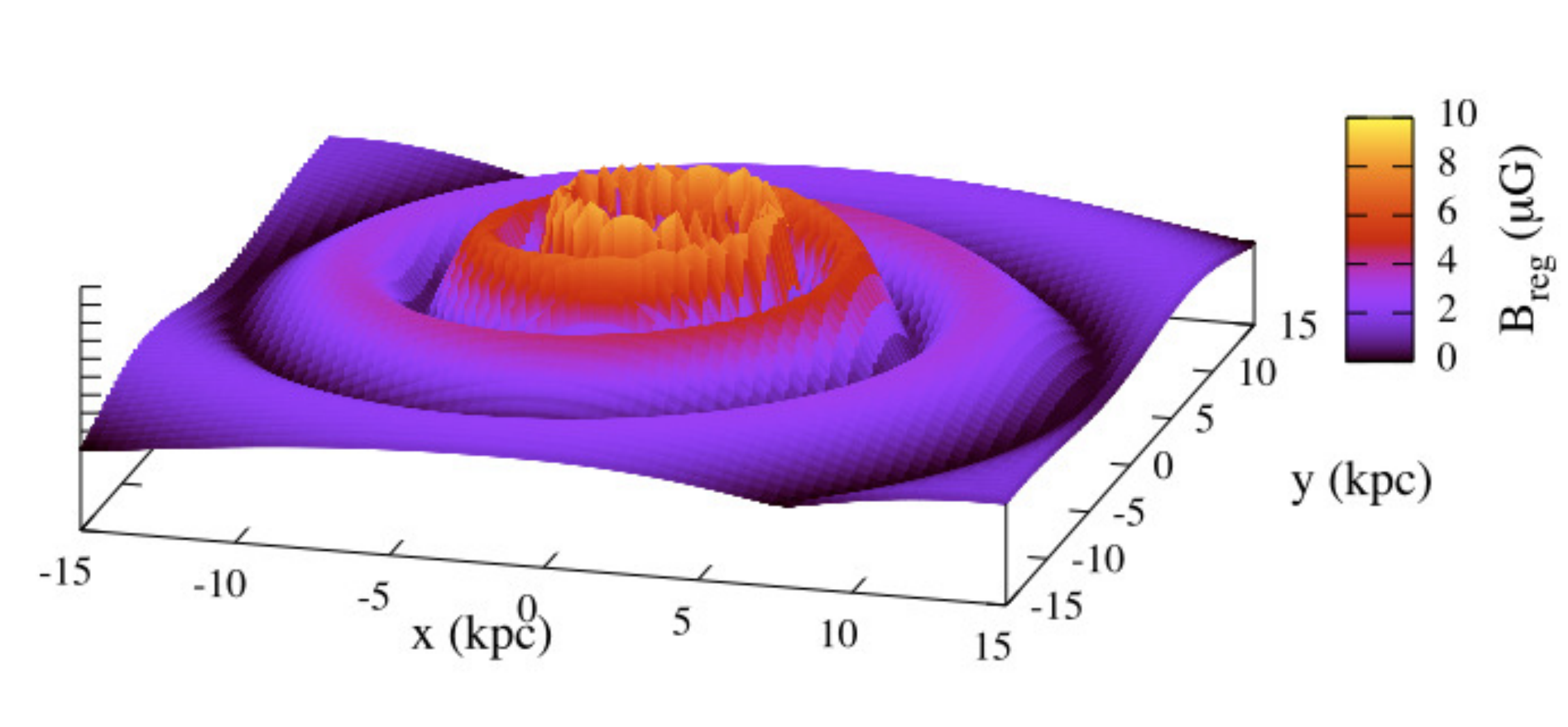}
     \includegraphics[width=7.cm]{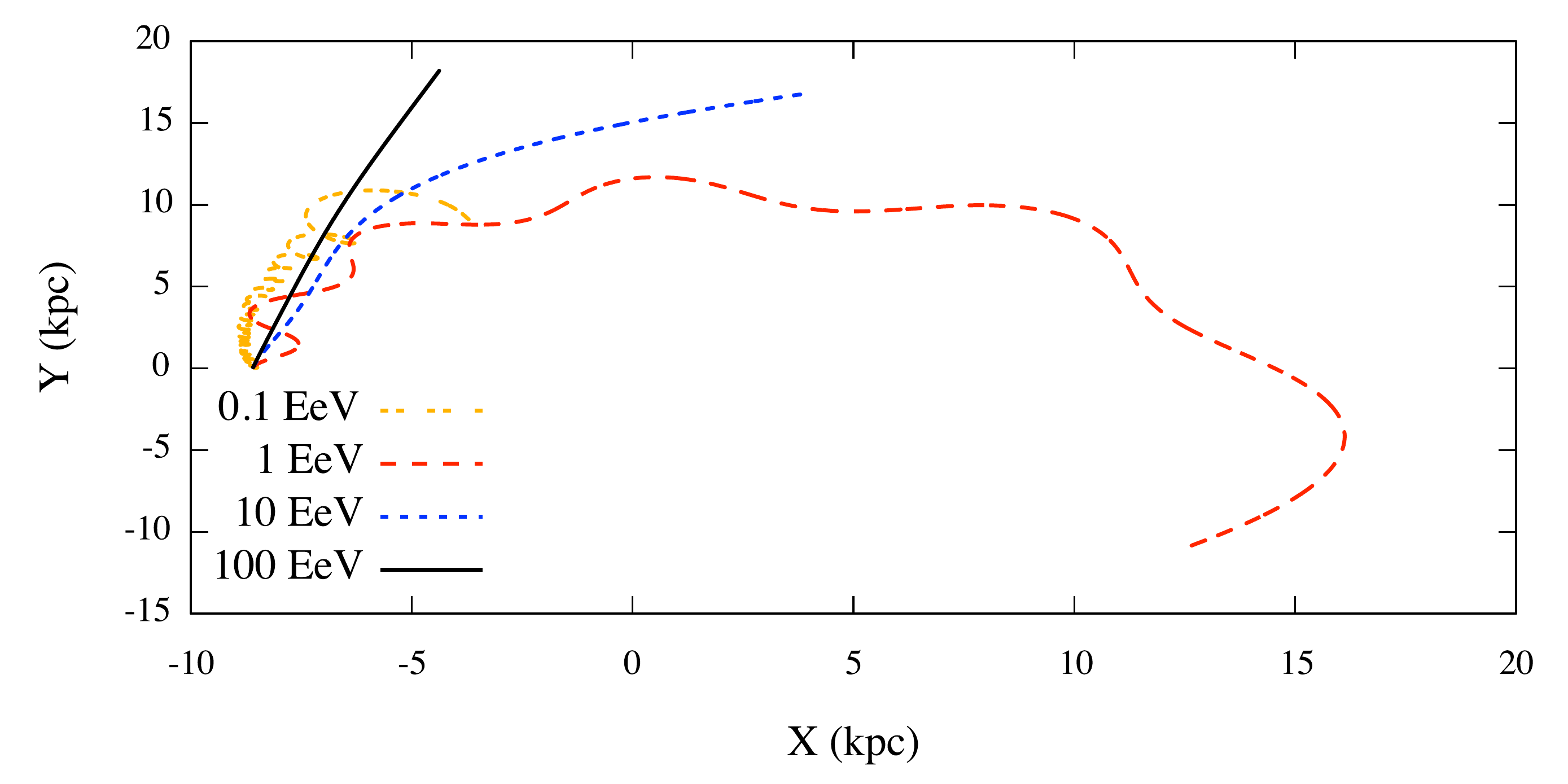}
    \caption{\textbf{Left:} HERMES simulation of the HMR model for the regular component of the magnetic field in our galaxy (at $z=0$), where the color indicates the intensity of the field. \textbf{Right:} Two-dimensional projection of the corresponding backtracked trajectories of UHECR for different values of the rigidity $E/Z$, ranging from $10^{17}$~eV (0.1~EeV) to $10^{20}$~eV (100~EeV)}
    \label{fig:mf-gal-bss3d}
  \end{center}
\end{figure}

In the left panel Fig.\,\ref{fig:mf-gal-bss3d} we show the HERMES simulation of the HMR model for the regular component of the magnetic field in our galaxy (at $z=0$). In the right panel of the same figure, the two-dimensional projection of the corresponding backtracked trajectories\footnotemark\footnotetext{A backtracked trajectory is the path traveled by the antiparticle, and it is obtained by substituting the charge $Z$ with the charge $-Z$ in the equations of motion.} of UHECR are shown for different values of the rigidity $E/Z$, ranging from $10^{17}$~eV (0.1~EeV) to $10^{20}$~eV (100~EeV). It is evident that at the lower energy particles tend to move along helical trajectories around the field lines, whereas for increasing energy particle tend to be less deflected.

\begin{figure}[!t]
  \begin{center} 
     \includegraphics[width=14cm]{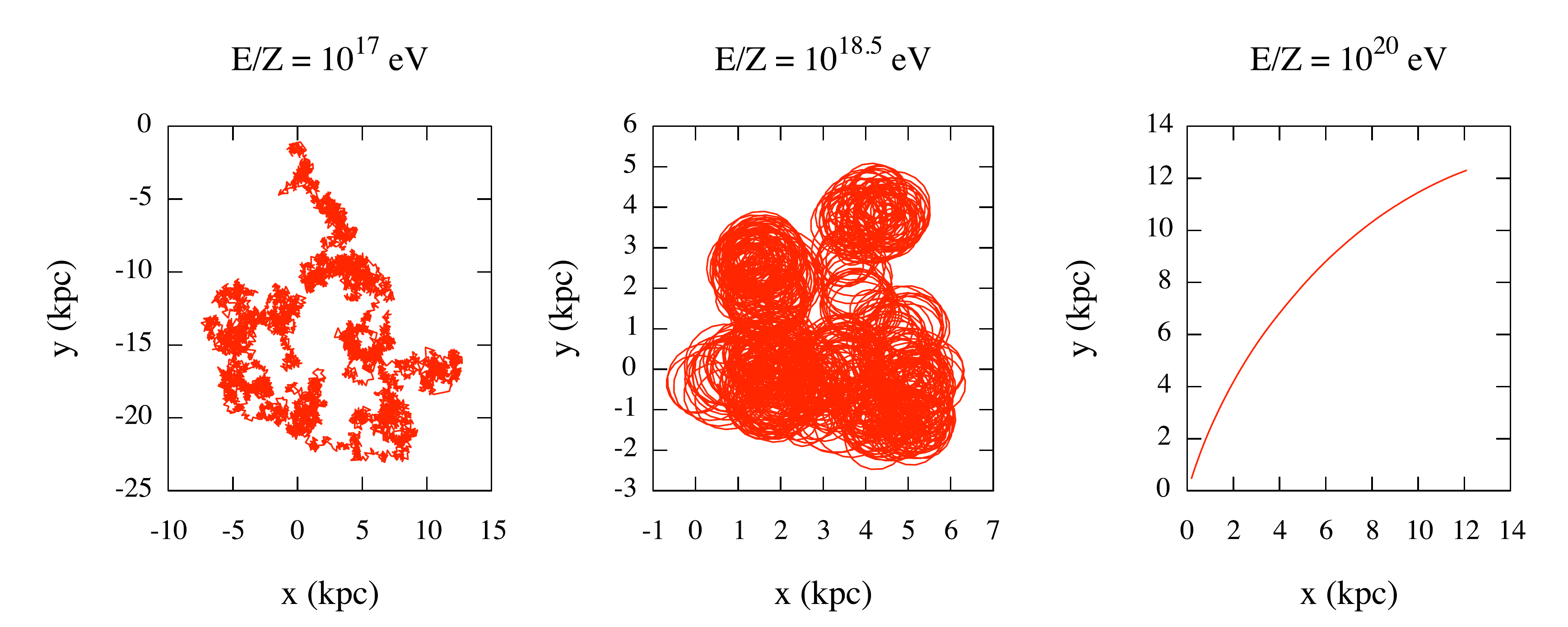}
    \caption{A random realization of nuclei trajectories in a uniform magnetic field ($B_{0}=3$~$\mu$G, parallel to the $z-$axis) plus a Kolmogorov 3D turbulent field ($\gamma=11/3$), for three different values of the ratio $E/Z$, namely $10^{17}$~eV, $10^{18.5}$~eV and $10^{20}$~eV. We have simulated the turbulent field ($\ell_{\text{max}}=100$~pc, $\langle B^{2}_{\text{irr}}\rangle=1$~$\mu$G, $\eta=0.1$) according to the Giacalone-Jokipii 3D isotropic approach.}
    \label{fig:mf-gal-turb-forwtrack}
  \end{center}
\end{figure}

For what concerns the irregular component of the GMF, as previously discussed, observations suggest a r.m.s. intensity of the order of the regular one, although no precise information is currently available. In Fig.\,\ref{fig:mf-gal-turb-forwtrack} we show the HERMES simulation of the trajectory (forward in time) of a particle with $E/Z$ ranging from $10^{17}$~eV to $10^{20}$~eV (left, middle and right panel, respectively), propagating in a magnetic field with an uniform component of intensity $B_{0}=3$~$\mu$G, parallel to the $z-$axis and a 3D turbulent component, characterized by maximum coherence length $\ell_{\text{max}}=100$~pc, r.m.s. strength $\langle B^{2}_{\text{irr}}\rangle=1$~$\mu$G and Kolmogorov index $\gamma=11/3$. The turbulence level is $\eta=0.1$. At the lowest energy the particle undergoes a brownian motion, being the Larmor radius of the order of turbulence scale $\ell_{\text{max}}$, whereas for increasing energy the particle only partially ``feels'' the turbulent component. At the highest energy the particle is subjected to the regular component only.

%%%%%%%%%%%%%%%%%%%%%%%%%%%%%%%%%%%%%%%%%%%
%%%%%%%%%%%%%%%%%%%%%%%%%%%%%%%%%%%%%%%%%%%

\section{Simulating the propagation of UHECR with HERMES: modeling interactions between UHECR and EBR photons}
\label{sec-modeling-interact}

We have shown the impact of magnetic fields on the propagation of UHE nuclei, without considering the energy-loss processes relevant for a complete study. This is the main subject of this section, where we show the impact of energy-loss processes on the propagation of UHE nuclei, photons and neutrinos. We will define the parameterizations we have chosen for the cross sections of the interactions between propagating UHECRs and photons of the background radiation and we will discuss all the relevant energy-loss processes included in our simulator as the adiabatic loss (due to the expansion of the Universe), the pair and photo-pion production, and, in the particular case of heavy nuclei, the photo-disintegration processes. The creation of secondary particles, produced by UHE nuclei undergoing pair and photo-pion production during their propagation, is also described: the development of the resulting UHECR cascade, including neutrinos and photons, will be briefly described to underline the complexity of simulating a realistic propagation. 

During their propagation, photons, neutrinos and nuclei $(A,Z)$ (electric charge, mass) with injection energy $E_{i}$, generally undergo interactions with background photons. UHECR that reach the Earth are therefore detected with a degraded energy $E_{f}<E_{i}$, depending on the type of interactions they were subjected to and on the distance between the source and the Earth. In HERMES, we describe the energy loss of non-stochastic processes in a unit interval of $z$ in terms of equations like
\begin{eqnarray}
\frac{1}{E}\frac{dE}{dz}=-\beta(z,E)\frac{dt}{dz},
\end{eqnarray}
where
\begin{eqnarray}
-\frac{dt}{dz}&=&\frac{1}{H_{0}(1+z)}\[ \Omega_{M}(1+z)^{3} + \Omega_{\Lambda} + (1-\Omega_{M}-\Omega_{\Lambda})(1+z)^{2} \]^{-\frac{1}{2}}
\end{eqnarray}
is the general metric element accounting for the cosmological expansion \cite{engel2001neutrinos, ave2005cosmogenic, stanev2009high}, and the involved cosmological parameters have been introduced in Sec.\,\ref{sec-cosmology}. The function $\beta(z,E)$ is related to the cooling rate of the UHE particle and it depends on the particular energy loss process considered. As we will see further in this section, $\beta(z,E)$ is proportional to the inverse of the mean free path and depends on the density of background photons and their energy, on the energy of the UHECR and on the cross section of the interaction under investigation. In the case of nuclei, it also depends on the nuclear mass and charge. Thus, the total energy loss rate is obtained by
\begin{eqnarray}
\label{def-energylosseq}
\frac{1}{E}\frac{dE}{dz} = -\frac{dt}{dz}\sum_{\text{process}}\beta_{\text{proc}}(z,E),
\end{eqnarray}
where the sum is extended to all interactions acting during the propagation. In HERMES, we include only those interactions which have a significant impact on the propagation of UHECR:
\begin{itemize}
\item \textbf{Adiabatic loss:} it is due to the expansion of the universe; it is considered for all nuclei with $A\geq 1$, photons and neutrinos;
\item \textbf{Pair production:} it involves the creation of a positron/electron pair; it is considered for all nuclei with $A\geq 1$ and photons;
\item \textbf{Photo-pion production:} it involves the creation of one or multiple pions; it is considered for all nuclei with $A\geq 1$;
\item \textbf{Photodisintegration:} it involves the fragmentation of the original nucleus, with the creation of lighter  nuclides (generally referred to as \emph{fragments}); it is considered for all nuclei with $A\geq 2$;
\item \textbf{Inverse Compton and synchrotron emission:} it is considered for photons and pairs which are part of the electromagnetic cascade generated by nuclei, and we refer to \cite{settimo2012eleca} for further details.
\end{itemize}

In the following we will take into account the interactions of nuclei with cosmic microwave background (CMB) and cosmic infrared/optical background (CIOB) radiations, by adopting the parameterization described in Sec.\,\ref{sec-ebr-parameterization} (see Fig.\,\ref{fig:ebl-model}) for the extragalactic background radiation. Eq.\,(\ref{def-energylosseq}) and mean free paths corresponding to the above interaction processes can be used to obtain an analytical approximation of the total energy loss. However, in order to obtain more realistic results, a Monte Carlo approach should be adopted for those processes where stochasticity is relevant, as in the case of photo-pion production and photodisintegration of heavier nuclei.

In the following, for the sake of simplicity, we will omit to specify that results shown in the following plots have been obtained from HERMES. At the end of this section we will describe the propagation of UHECR with no regards of magnetic fields: such an approach is generally known as ``1D propagation''.

\subsection{Adiabatic loss}

In order to take into account the energy loss due to the expansion of the universe, we use 
\begin{eqnarray}
\label{def-betarsh}
\beta_{\text{rsh}}(z)&=& H_{0}\[\Omega_{M}(1+z)^{3} + \Omega_{\Lambda} + (1-\Omega_{M}-\Omega_{\Lambda})(1+z)^{2} \]^{\frac{1}{2}}
\end{eqnarray}
for the adiabatic term, as previously explained in Sec.\,\ref{sec-cosmology}.

\subsection{Cross section of $A\gamma$ nuclear interactions}
\label{sec-cross-A}

The probability of UHE protons to interact with background photons rapidly increases with proton energy. If $E$ and $\epsilon$ are the energies of the UHE proton and the photon in the observer rest frame, respectively, the interaction is equivalent to a collision with a high energy photon with energy $\epsilon'=\Gamma\epsilon(1-\cos\theta)$, being $\theta$ the collision angle. For instance, when the energy $\epsilon'$ equals at least the pion mass $m_{\pi}c^{2}\approx 140$ MeV, the proton undergoes photo-meson production and loses energy. Such a process is known as Greisen-Zatsepin-Kuzmin effect and dominates above $50-60$ EeV \cite{greisen1966end,zatsepin1966upper}. The two main channels for $p+\gamma_{\text{EBR}}$ interaction, involving the resonance $\Delta(1232 \text{ MeV})$ close to the threshold energy, are $\Delta(1232~\text{MeV})\lto p+\pi^{0}$ and $\Delta(1232~\text{MeV})\lto n+\pi^{+}$ (with the consequent channel $n\lto p +e^{-}+\bar{\nu}_{e}$). At higher energies, heavier resonances and multi-pion production channels are likely. Just above the threshold, baryonic resonances dominate and protons are subjected to photo-meson production, mainly through the $\Delta(1232)$-baryon resonance, whereas heavier resonances (up to $\Delta(1950)$-baryon) play a more marginal role. We parameterize the cross-section for baryonic resonances by
\begin{eqnarray}
\sigma_{\text{BR}}(\epsilon)&=& \sum_{i=1}^{4}\sigma_{i}\sigma_{L}(\epsilon;\epsilon_{i},\Gamma_{i})\nonumber
\end{eqnarray}
where $\sigma_{L}$ is the Lorentzian function, ($\epsilon_{i}$ (GeV), $\Gamma_{i}$ (GeV), $\sigma_{i}$ ($\mu$b)) $=(0.34, 0.17, 351)$, $(0.75, 0.50,159)$, $(1.00, 0.60, 21)$ and $(1.50, 0.80, 26)$ for $i=1,2,3$ and 4, respectively. For all other processes participating in photo-meson production, including multipions (MP) or direct particle production involving $\pi$, $\eta$, $\Delta$, $\rho$, $\omega$ and strange-particle channels (RP), we use Rachen's parameterizations \cite{rachen1996thesis}. In the following we will use the abbreviation ``BR'' to refer to baryonic resonances, direct particle and multi-pion production, where not specified otherwise.

As in the case of protons, the probability of heavier UHE nuclei to interact with background photons rapidly increases with nucleus energy. The processes involved in such interactions are the same that we have previously described in the case of protons, namely pair and photo-pion production. However, in the case of heavy nuclei we have to take also into account the photo-disintegration (or photo-erosion) process
\begin{eqnarray}
^{A}_{Z}X+\gamma \lto ^{A'}_{Z'}Y + m\alpha + [(Z-Z')-2m]p + [(A-A')-(Z-Z')-2m]n,
\end{eqnarray}
resulting in the emission of subatomic particles, with the creation of lighter nuclides. Here, $m$ is the multiplicity of $\alpha$ particles, $p$ indicates the proton and $n$ the neutron. In general, in order to describe the changes in abundance of the heavy nuclei as a result of the interaction of the UHECR with the background radiation, a nuclear reaction network including all interactions of interest should be used. Such a network is described by a system of coupled differential equations corresponding to all the reactions affecting each nucleus, i.e. mainly photo-disintegrations and $\beta-$decays. Such an approach has been recently proposed, and adopted in many successive works, in Ref.\,\cite{allard2005uhe} for the study of UHE nuclei propagation by using up to date measurements of cross sections \cite{khan2005photodisintegration}. Instead of direct measurements, other recent works related to this topic \cite{kampert2009propagation,ahlers2010analytic,ahlers2011cosmogenic} make use of TALYS \cite{koning2005talys,TALYSweb}, a software for the most likely simulation of nuclear reactions.

We adopt the simplest approach to the treatment of the photo-disintegration channels, by following the chain of stable nuclei (stability chain), as suggested for the first time by Puget, Stecker and Bredekamp (PSB) \cite{puget1976photonuclear}. The relative contribution of all decay channels corresponding to nuclei with $A\leq 56$ are taken from Ref.\,\cite{puget1976photonuclear} and \cite{stecker1999photodisintegration}. However, in order to produce more realistic simulations of the photo-disintegration process, we have obtained from TALYS reactions the branching ratios associated to the most relevant exclusive channels, including one nucleon, two nucleons and multi-nucleons emission on CMB and CIOB, similarly to recent studies \cite{kampert2009propagation,ahlers2010analytic}. Hence, in HERMES, we have included different models for the photo-disintegration of nuclei, with cross sections corresponding to: i) the PSB Gaussian approximation; ii) the Rachen's parameterizations; iii) the TALYS reactions. See the corresponding referenced works for further details.

\begin{figure}[!t]
  \begin{center} 
       \includegraphics[width=12cm]{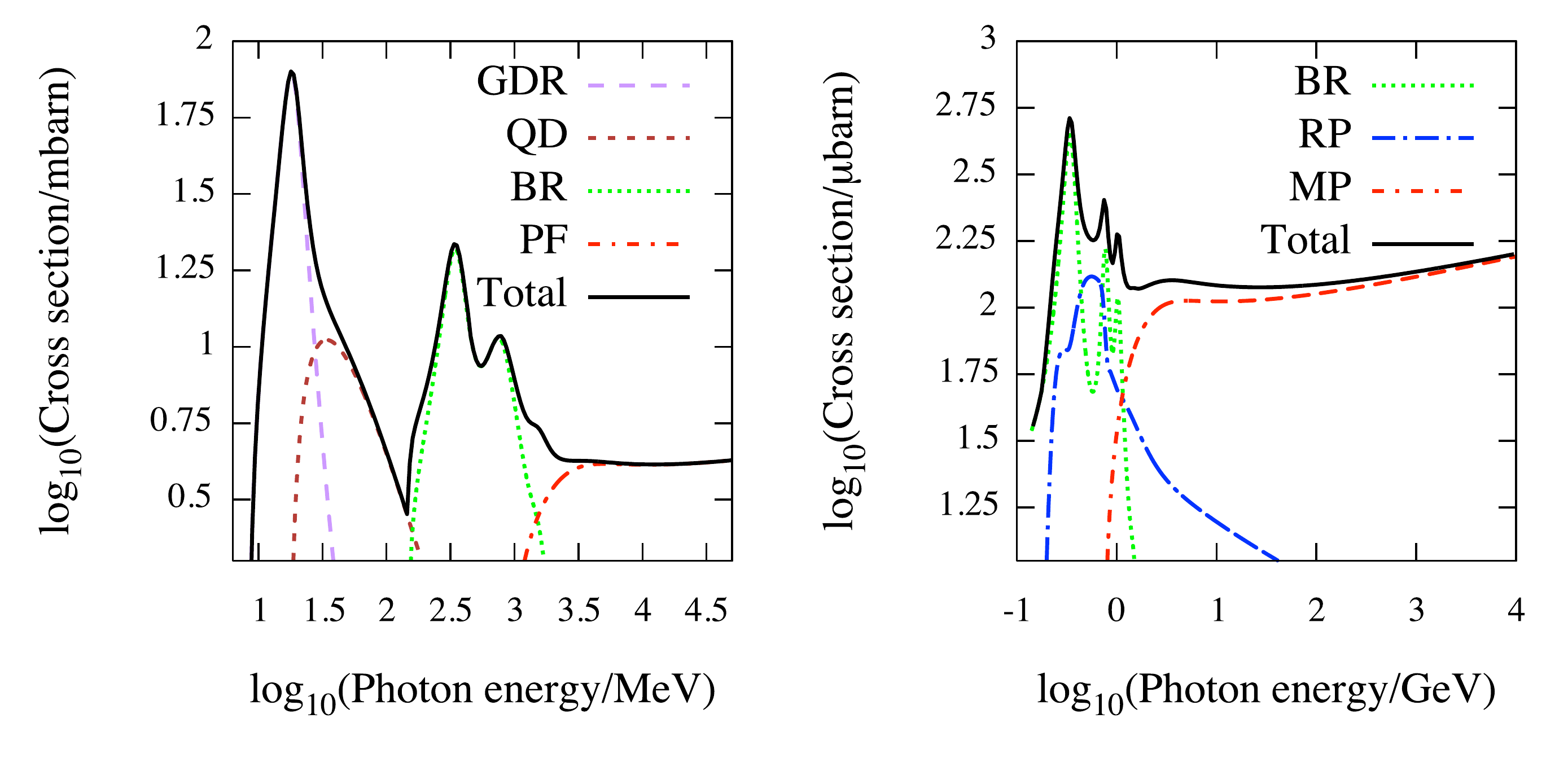}
    \caption{Comparison between the total cross section for Fe$\gamma_{\text{EBR}}$ (left panel) and $p\gamma_{\text{EBR}}$ (right panel) interactions as a function of the background photon energy $\epsilon'$ in the nucleus rest frame, obtained from our HERMES, following Rachen's parameterizations\,\cite{rachen1996thesis,stanev2007icrc}. Different contributions due to baryonic resonances (BR), direct particle (RP) and multi-pion (MP) production are shown as reported in Refs. (left panel) and as obtained from HERMES, following Rachen's parameterizations (right panel).}
    \label{fig:prop-cross-p-fe}
  \end{center}
\end{figure}

In Fig.\,\ref{fig:prop-cross-p-fe} we show the comparison between the total cross sections estimated for iron (left panel) and proton (right panel) nuclei, together with the contribution of each single process separately.

For the sake of completeness, it is worth remarking that in HERMES the inclusion of some additional processes, not depending on the background radiation, are currently under development:
\begin{eqnarray}
^{A}_{Z}X\lto ^{A}_{Z+1}Y+e^{-}+\bar{\nu}_{e}&\quad&\beta^{-}-\text{decay},\nonumber\\
^{A}_{Z}X\lto ^{A}_{Z-1}Y+e^{+}+\nu_{e}&\quad&\beta^{+}-\text{decay},\nonumber\\
^{A}_{Z}X+e^{-}\lto ^{A}_{Z-1}Y+\nu_{e}&\quad&\text{electron capture}.\nonumber
\end{eqnarray}

\subsection{Interaction lengths for $A\gamma$ interactions}

The adiabatic loss is considered during the whole propagation as a continuous energy loss process. Instead, the interaction length (or, equivalently, mean free path) corresponding to different processes is used as an input to the Monte Carlo algorithm to randomly sample the next point where the nucleus will undergo one of the interactions described at the end of Sec.\,\ref{sec-modeling-interact}. Such interactions are treated as competitive processes, except the pair production which is treated as a continuous energy loss in the current version of HERMES (see further in the text). Therefore, the estimation of the interaction lengths is fundamental and allows to simulate the production of secondary UHECR (lighter nuclei from photo-disintegration, neutrinos and photons cascades). The interaction length is given by
\begin{eqnarray}
\label{def-lambda}
\mathcal{\lambda}^{-1}_{A}(z,E)=\mathcal{E}(z)\frac{c}{2\Gamma_{A}^{2}}\int_{\epsilon_{\text{thr}}/2\Gamma_{A}}^{\epsilon_{\text{max}}}d\epsilon\frac{n(\epsilon)}{\epsilon^{2}}\int_{\epsilon_{\text{thr}}}^{2\Gamma_{A}\epsilon}d\epsilon'\epsilon'\sigma(\epsilon')
\end{eqnarray}
where $\Gamma_{A}=(1+z)\frac{E}{Am_{p}c^{2}}$ is the Lorentz factor of the nucleus at redshift $z$, $\epsilon_{\text{thr}}$ is the energy threshold of the considered process in the nucleus rest frame, $n(\epsilon)$ is the density of background photons with energy $\epsilon$ in the observer's rest frame, $\epsilon'$ is the energy of the photon in the nucleus rest frame and $\mathcal{E}(z)$ is the evolution function of the ambient photon field. It is straightforward to show that $\lambda_{A}(z,E)=(1+z)^{-3}\lambda_{A}[z=0,(1+z)E]$ when the CMB is considered \cite{stanev2000propagation,berezinsky2006astrophysical}, whereas for other background radiations a more complicated evolution should be used. By following Stanev et al \cite{stanev2000propagation}, we define the average energy loss length by 
\begin{eqnarray}
\chi_{\text{loss}}(z,E)=\frac{E}{dE/dz}=\frac{\lambda_{A}(z,E)}{\kappa(E)},
\end{eqnarray}
where $\kappa(E)=\langle\Delta E\rangle/E$ is the mean inelasticity, i.e. the average fraction of energy lost by the nucleus because of the interaction. The inelasticity for pair production is $\kappa\approx 2m_{e}/(Am_{p})$ (being $m_{e}$ and $m_{p}$ the masses of electron and proton, respectively), i.e. around $10^{-3}$ in the case of protons, and even smaller for heavier nuclei. Conversely, for photo-pion production by protons, the inelasticity ranges from 0.2 to 0.5, depending on the energy. 

In the case of heavy nuclei, the differences in the cross section (with respect to the case of protons) are reflected in the interaction length. In Fig.\,\ref{fig:prop-cross-p-fe}, the available channels above the threshold for single pion production ($\epsilon'\approx 145$~MeV) involve baryonic resonances and direct particle production, with multi-pion production playing a significant role at the highest energies ($\epsilon'>700$~MeV). In the case of iron, the additional channels due to photo-disintegration process are evident at lower energies ($1<\epsilon'<150$~MeV). The energy loss due to the pair production, Eq.\,(\ref{def-betapair}), and to the adiabatic loss, Eq.\,(\ref{def-betarsh}), occurs in any case, with significant contributions only in a small range of energies. We treat the photo-pion production similarly to the case of protons by using Eq.\,(\ref{def-lambda}) and the $\Delta-$baryon decay channels. The energy loss equation, defined by Eq.\,(\ref{def-energylosseq}), still applies but coupled to the nuclear mass loss rate
\begin{eqnarray}
\frac{1}{A}\frac{dA}{dz}=-\frac{dt}{dz}\beta_{\text{dis,eff}}(z,E;Z,A),
\end{eqnarray}
leading to
\begin{eqnarray}
\frac{1}{E}\frac{dE}{dz}=\frac{1}{\Gamma}\frac{d\Gamma}{dz}+\frac{1}{A}\frac{dA}{dz}.
\end{eqnarray}
An analytic approach for the estimation of the spectra at Earth, based on the numerical integration of such an equation, has been recently reported in \cite{aloisio2008analytic}.

In the rest frame of the nucleus, the pair production process $A+\gamma_{\text{EBR}}\lto A+e^{+}+e^{-}$ occurs at the threshold energy $2m_{e}c^{2}\approx 1$~MeV and it plays an important role only when CMB is considered, the CIOB participating marginally \cite{puget1976photonuclear}. We can treat the process as a continuous energy loss, because the loss per interaction is very small. In HERMES, the energy loss accounting for the pair production, due to the Bethe-Heitler interaction with ambient photons with density $n(\epsilon)$, is given by \cite{blumenthal1970energy}
\begin{eqnarray}
\label{def-betapair}
\beta_{e^{\pm}}(z,E;Z,A)\simeq S(Z)\frac{\alpha^{3}Z^{2}A^{2}}{4\pi^{2}\hbar}\frac{m_{e}^{2}m_{p}^{2}}{E^{3}}\int_{2}^{\infty}d\xi\frac{\varphi(\xi)}{\exp\[\frac{m_{e}Am_{p}}{2E(1+z)k_{B}T_{0}}\xi\]-1},
\end{eqnarray}
that is similar to the parameterization adopted in \cite{cuoco2006footprint}, where the auxiliary function $\varphi(\xi)$ is obtained from \cite{chodorowski1992reaction} and masses are in units of eV/$c^{2}$. However, there is no parameterization in the case of CIOB and, in our code, we estimate the corresponding energy loss rate by using Eq.\,(\ref{def-lambda}). In Eq.\,(\ref{def-betapair}), $\gamma\approx E/(Am_{p}c^{2})$ is the Lorentz factor of the nucleus, $m_{e}$ is the electron mass, $\alpha=e^{2}/\hbar c$ is the fine-structure constant, $r_{e}=e^{2}/m_{e}c^{2}$ is the classical electron radius, $T_{0}=2.725$~K and $k_{B}$ is the Boltzmann constant. The factor $S(Z)$ is a correction term to agree with experimental data for nuclei with $Z>1$ \cite{rachen1996thesis}, even if it has been pointed out that Coulomb corrections to the Born approximation have a negligible effect on the pair production loss rate of ultra-relativistic heavy nuclei as $^{56}$Fe \cite{stecker1999photodisintegration}. 

Concerning the photo-pion production process (see Sec.\,\ref{sec-cross-A}), in the particular case of protons propagating in the CMB, Eq.\,(\ref{def-lambda}) at present time reduces to
\begin{eqnarray}
\beta_{\pi}(E)=-\frac{k_{B}T_{0}}{2\pi^{2}\hbar}\frac{m^{2}_{p}}{E^{2}}\int_{0}^{\infty}d\epsilon \kappa(\epsilon)\sigma(\epsilon)\epsilon\times\ln\[1-\exp\(-\frac{m_{p}}{2Ek_{B}T_{0}}\epsilon\)\],
\end{eqnarray}
where $m_{p}$ is the proton mass in units of eV$/c^{2}$, $\sigma(\epsilon)$ is the cross-section for pion production in terms of the photon energy $\epsilon$ and $\kappa(\epsilon)$ is the inelasticity factor. In order to avoid further numerical integrations, we parameterize the contribution of this term as in Ref. \cite{cuoco2006footprint} by
\begin{eqnarray}
\beta_{\pi}(z,E;1,1)\simeq\left\{
\begin{array}{ll}
A_{\pi}(1+z)^{3}\exp\[\frac{B_{\pi}}{E(1+z)}\] & E\leq E_{\text{match}}(z)\\
C_{\pi}(1+z)^{3} & E> E_{\text{match}}(z)
\end{array}\right..
\end{eqnarray}
The function $E_{\text{match}}(z)=6.86 e^{-0.807z}\times 10^{20}$ eV ensures the continuity of the function $\beta_{\pi}(z,E;1,1)$ and $\{A_{\pi},B_{\pi},C_{\pi}\}=\{3.66\times10^{-8}\text{yr}^{-1},2.87\times10^{20}\text{eV},2.42\times10^{-8}\text{yr}^{-1}\}$ are taken from Ref. \cite{anchordoqui1997effect}. 

We treat the case of neutron in a similar way, by considering the additional process of the $\beta-$decay. The neutron decay rate is given by $m_{N}/(\tau_{n}E)$, with $\tau\simeq 888.6$~s the laboratory lifetime, providing a range of propagation $\lambda_{\beta}=\tau_{n}\frac{E}{m_{N}}\simeq 0.9\(E/10^{20}~\text{eV}\)~\text{Mpc}$, which becomes competitive with photo-pion production only at the highest energy, above $10^{21}$~eV.

\begin{figure}[!t]
  \begin{center} 
       \includegraphics[width=7.cm]{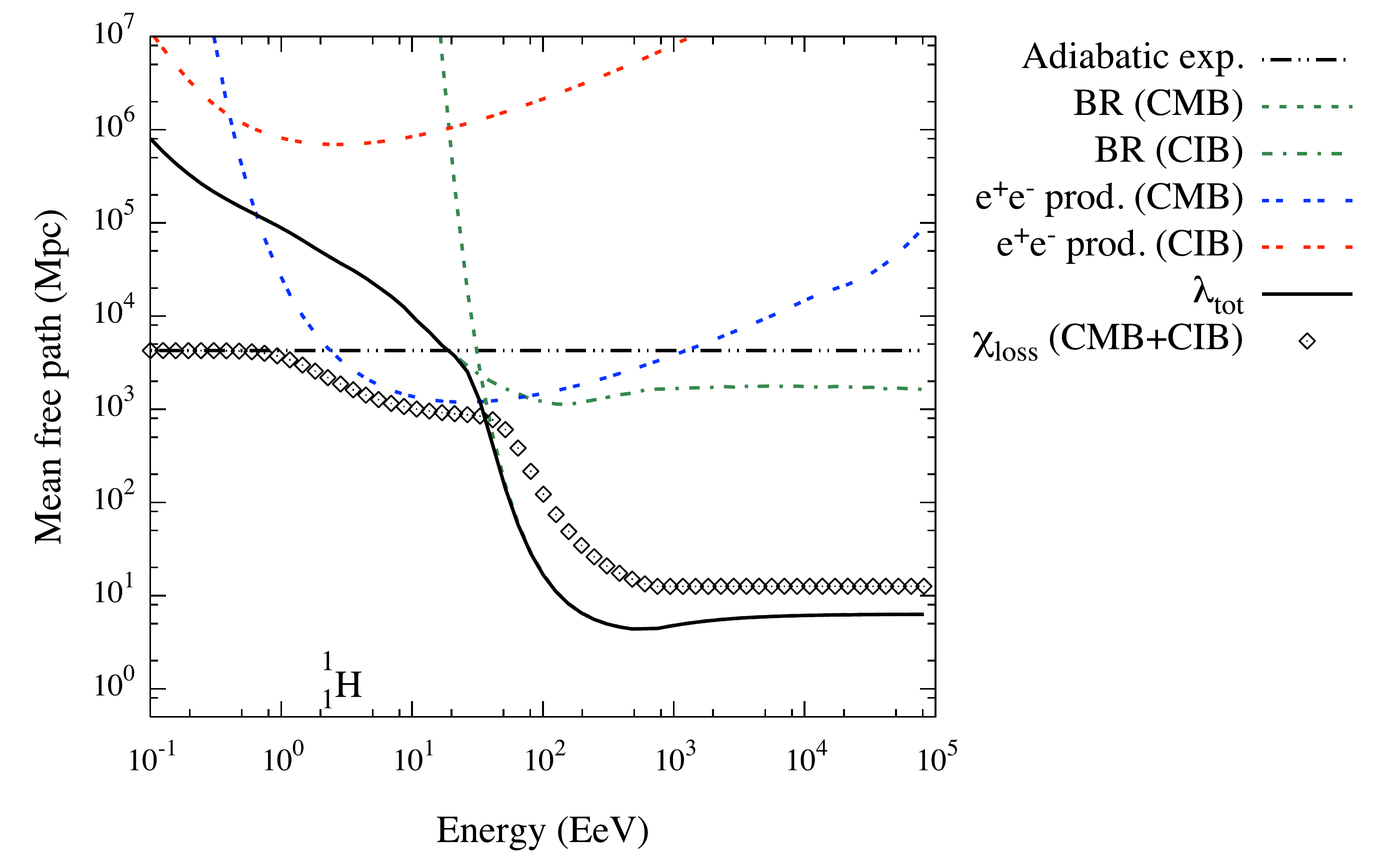}
       \includegraphics[width=7.cm]{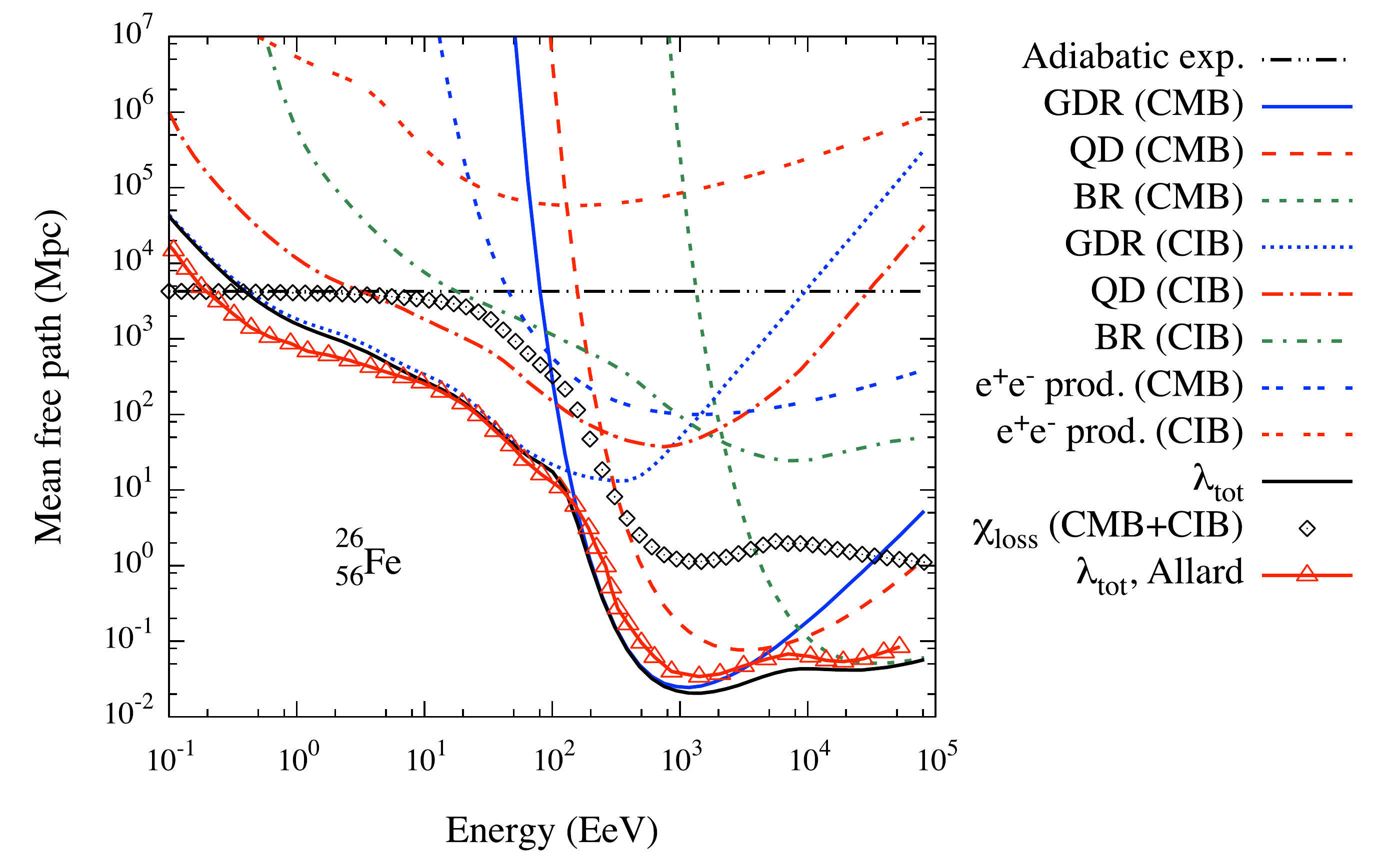}
    \caption{Our estimation of the mean free path $\lambda$ at $z=0$ as a function of the energy of the nucleus in the observer rest frame. The contributions due to different processes (adiabatic, pair and photo-pion production, as well as giant dipole resonance and quasi-deuteron effect for photo-disintegration) in CMB and CIOB are shown separately. The total interaction length $\lambda_{\text{tot}}$ and the total energy loss length $\chi_{\text{loss}}$ are shown as well in the case of proton (left panel) and Iron $^{56}_{26}Fe$ (right panel) nuclei: for the latter, the $\lambda_{\text{tot}}$ estimated by Allard et al \cite{allard2006cosmogenic,allard2009propagation} is reported for reference.}
    \label{fig:prop-lambda-pfe}
  \end{center}
\end{figure}

In Fig.\,\ref{fig:prop-lambda-pfe} we show the interaction length $\lambda$ of proton (left panel) and iron (right panel) nuclei, in the CIOB and the CMB, for each process separately and for all processes together, as well as the energy loss length $\chi_{\text{loss}}$, as a function of the energy $E$ of the nucleus in the observer rest frame at present time ($z=0$). 

In the case of proton, it is evident that the pair production on CIOB is negligible with respect to other processes, because occurring on time scales larger than the adiabatic expansion, for all energies above $10^{18}$~eV. A similar argument applies for the photo-meson production in the CIOB, which, below $10^{20}$~eV, contributes less than pair production in the CMB, whereas above $10^{20}$~eV the production of pions in the CMB dominates up to the highest energy. In the energy interval between $2\times10^{18}$~eV and $\sim5\times10^{19}$~eV, the main energy loss process is the pair production in the CMB. The obtained results are in perfect agreement with recent literature \cite{berezinsky2006astrophysical,harari2006ultrahigh,stanev2009propagation,kotera2011astrophysics}, with small differences related to the different CIOB adopted.

In the case of iron, the figure shows that the main energy loss below $10^{19}$~eV is due to the adiabatic expansion of the universe, whereas photo-disintegration process through the giant dipole resonance dominates above $10^{19}$~eV and photo-meson production becomes dominant above $10^{22}$~eV. 

\begin{figure}[!t]
  \begin{center} 
       \includegraphics[width=12cm]{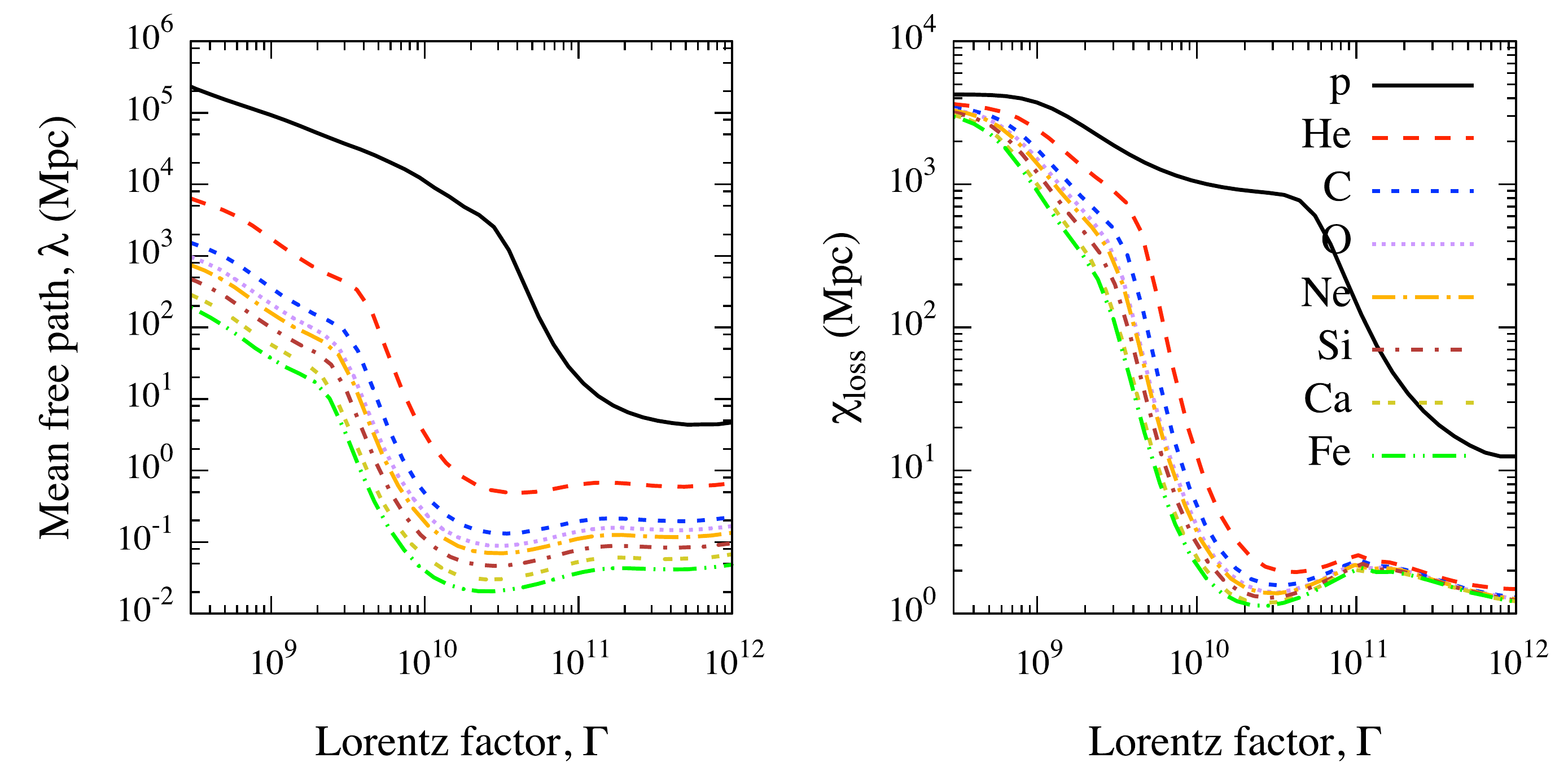}
    \caption{\textbf{Left:} Total mean free path $\lambda_{\text{tot}}$ in CMB and CIOB for several nuclei at $z=0$, from proton ($A=1$) to iron ($A=56$), as a function of the Lorentz factor $\Gamma$. \textbf{Right:} Same as in the left panel, but for the total energy loss length $\chi_{\text{loss}}$.}
    \label{fig:prop-xloss-nuclei}
  \end{center}

%  \begin{center} 
%       \includegraphics[width=12cm]{PROP_xloss_evolution_p_fe}
%    \caption{Evolution of the total energy loss length $\chi_{\text{loss}}$ in CMB and CIOB as a function of the energy of the nucleus in the observer rest frame, for different values of the redshift. \textbf{Left:} Case of proton. \textbf{Right:} Case of iron.}
%    \label{fig:prop-xloss-nuclei-evol}
%  \end{center}
\end{figure}

The estimation of the total interaction (left panel) and energy loss (right panel) lengths at $z=0$ obtained with HERMES are also shown in Fig.\,\ref{fig:prop-xloss-nuclei} as a function of the Lorentz factor $\Gamma$, for several nuclei, from proton to iron. Both quantities decrease for increasing nuclear mass and for any value of the energy, although the energy loss length tends to become constant for all nuclei above $\Gamma=10^{11}$, approximately the value where baryonic resonances occur. %In Fig.\,\ref{fig:prop-xloss-nuclei-evol} we also show the evolution with redshift of the total energy loss length as a function of the energy, for a proton (left panel) and for an iron (right panel) nucleus, for different values of the redshift, ranging from $z=0$ to $z=4$.

%%%%%%%%%%%%%%%%%%%%%%%%%%%%%%%%%%%%%%%%%%%%
 
\subsection{Propagation of secondary neutrinos and photons}

We have discussed the production of electron/positron pairs and of secondary pions. Produced UHE photons and pairs interact with the extragalactic background photons, participating to the electromagnetic cascade generated by the primary nucleus. Conversely, in the case of photo-meson production, pions have small lifetime, of the order of $10^{-16}$~s for $\pi^{0}$ and $10^{-8}$~s for $\pi^{\pm}$: thus, we neglect their propagation because they quickly decay to new secondary particles, which can decay to other particles (as in the case of secondary muons) generating a cascade of electrons, positrons, photons and neutrinos. In HERMES, we consider all the main decay channels involving the production of a single pion and we include the $\beta-$decay of neutrons.

Additionally, channels with multi-pion production are present. As shown in \cite{mucke124sophia}, close to the threshold and for $\epsilon'<1$~GeV, the dominating processes involve single pion production only, whereas at the highest energies channels with two or three pions are available. The inclusion in HERMES of such channels is currently under development.

The propagation of UHE photons produced by neutral pions, and the consequent pairs, are performed with EleCa and its description is beyond the scope of the present work. We refer to \cite{settimo2012eleca} for further details. 

Neutrinos, produced by the decay of charged pions and $\beta-$decay of neutrons, are chargeless particles with negligible mass, undergoing interactions only through the weak nuclear force (and gravity, if they are considered massive particles). Because of such features, neutrinos are likely to traverse the extragalactic space, even for cosmic distances, without interacting with background photons or interstellar medium, and without being deflected by magnetic fields: characteristics that makes neutrinos the ideal candidates for particle astronomy. On the other hand, the flux of cosmogenic neutrinos is relatively small if compared to the flux of charged particles, at the highest energy. Propagation and energy loss of neutrinos, can be easily described by energy loss equation (\ref{def-energylosseq}), considering only the adiabatic energy loss rate defined by Eq.\,(\ref{def-betarsh}).

%%%%%%%%%%%%%%%%%%%%%%%%%%%%%%%%%%%%%%%%%%%
%%%%%%%%%%%%%%%%%%%%%%%%%%%%%%%%%%%%%%%%%%%

\section{Applications}

In this section we briefly discuss some applications to show the potentiality of HERMES for studying UHECR, including the comparison between results obtained with HERMES and those either from other propagation codes available in the UHECR community or from observation. 

\begin{figure}[!t]
\centering
%\subfigure[Surviving probability]
%   {
   \includegraphics[width=11cm]{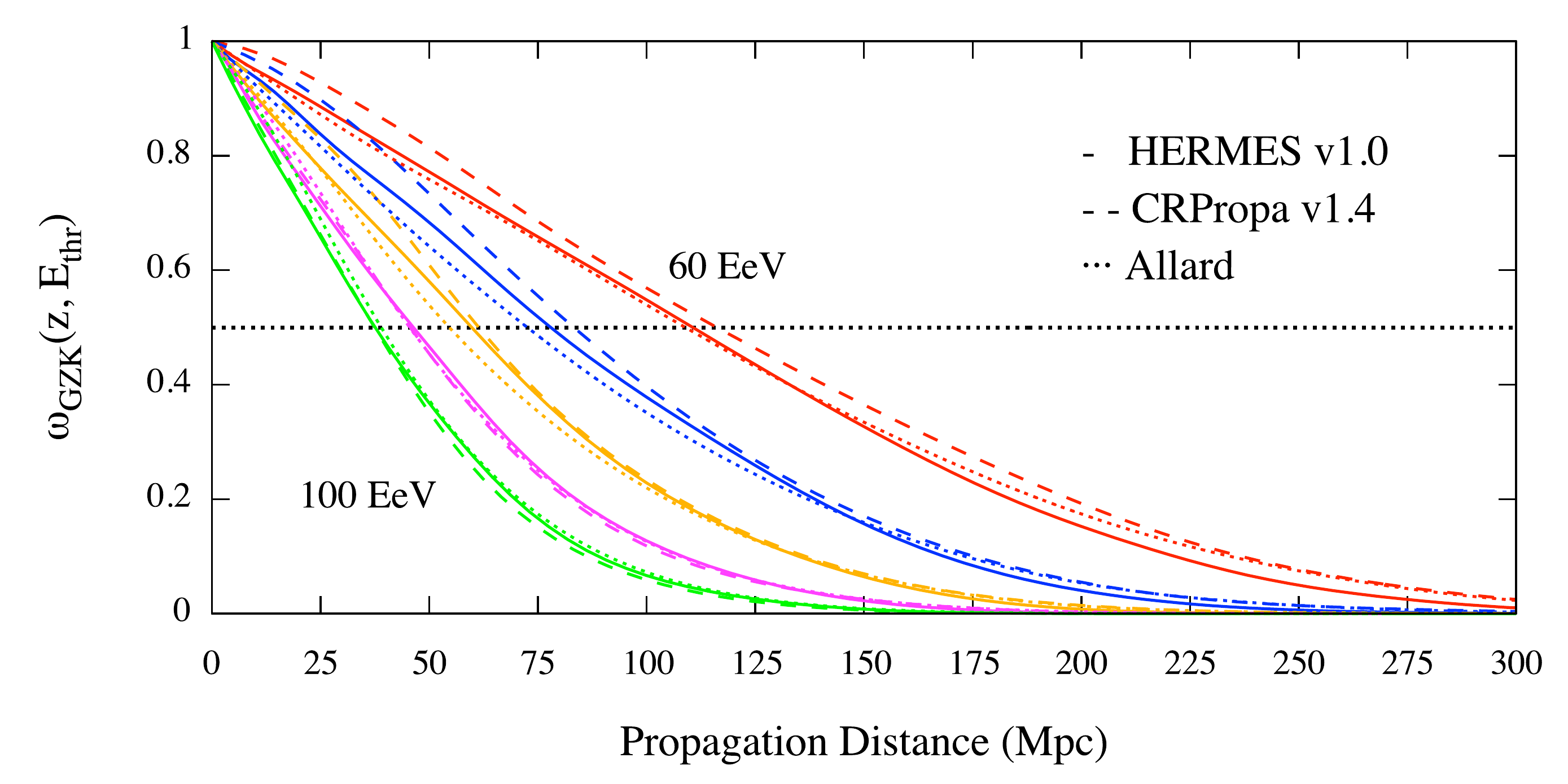}
   \caption{Surviving probability of protons (see the text) produced by an homogenous distribution of sources within 300~Mpc, as a function of the propagation distance and for different energy threshold $E_{\text{thr}}$ at Earth. A power-law injection with energy cutoff $10^{21}$~eV and spectral index $2.4$ is used. We show the result of our simulations performed with HERMES (solid line), compared to those obtained with CRPropa~v1.4 (dashed line) and by Allard (dotted line).}
   \label{fig-hermes-comp2}
\end{figure}

First, we investigate the surviving probability $\omega_{\text{GZK}}(z,E_{\text{thr}})$ of protons, i.e. the probability that a proton produced by a sources at redshift $z$ could reach the Earth with an energy above a given threshold. We consider an homogenous distribution of equal-intrinsic-luminosity sources in the nearby Universe, up to $\approx300$~Mpc: each source emits protons following a power-law injection spectrum with spectral index $2.4$ and energy cutoff $10^{21}$~eV. Hence, we estimate $\omega_{\text{GZK}}(z,E_{\text{thr}})$ for different energy threshold $E_{\text{thr}}$ at Earth, ranging from 60~EeV to 100~EeV. The result is shown in Fig.\,\ref{fig-hermes-comp2}, where a comparison between HERMES, CRPropa~v1.4~\cite{armengaud2007crpropa} and D. Allard \emph{et al} \cite{allard2005uhe}, are reported. The resulting curves are in good agreement, putting in evidence the goodness of our simulator.

\begin{figure}[!t]
	\centering
	  \includegraphics[width=11.0cm]{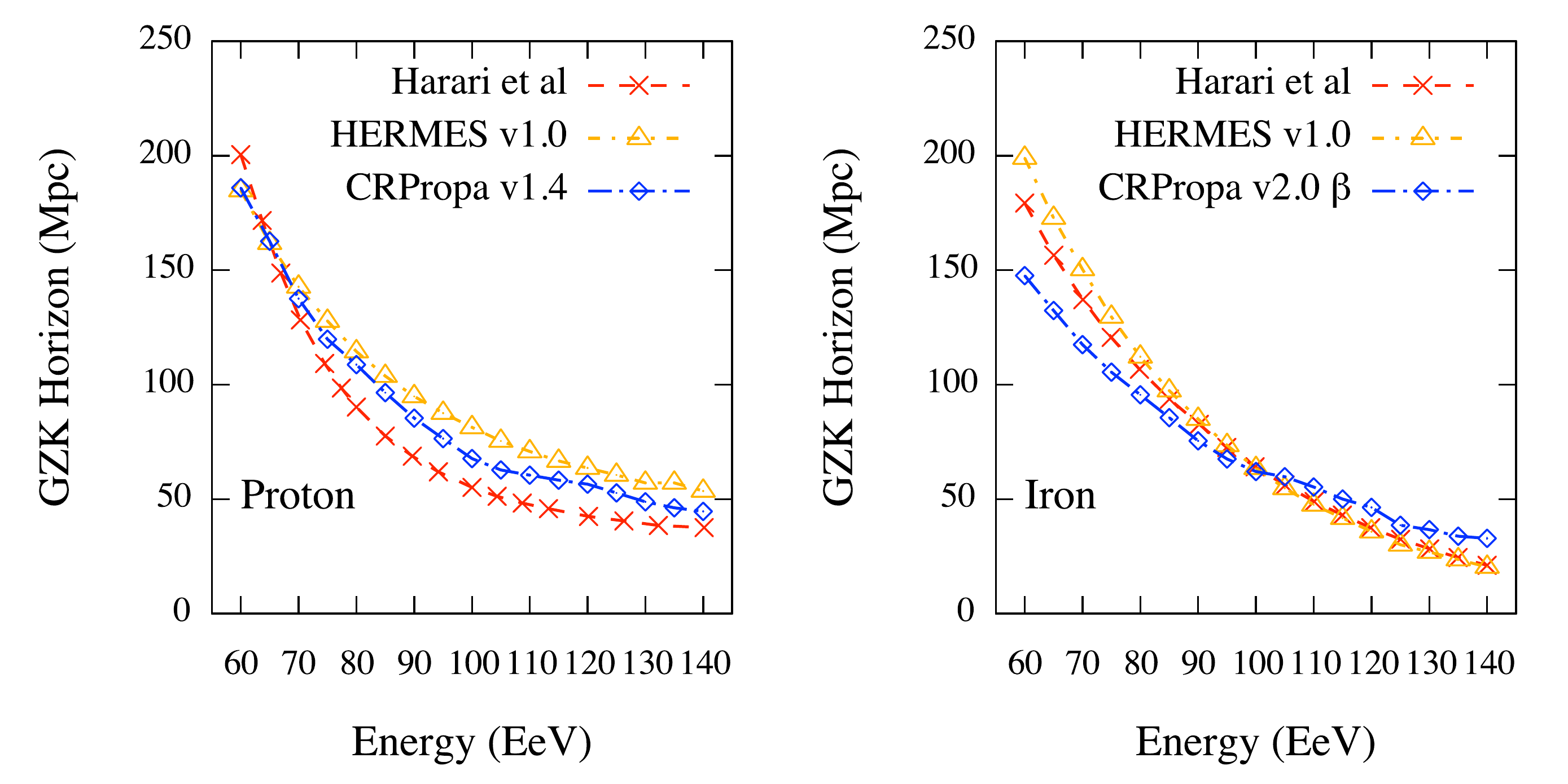}
	\caption{GZK horizon estimated in the case of protons (left panel) and iron nuclei (right panel) injected with spectrum $E^{-2.7}$, as a function of the energy threshold at Earth. Results from CRPropa and Harari \emph{et al} \cite{harari2006ultrahigh} are shown for reference.}
\label{fig-hermes-comp4}
\end{figure}

%   }
% \hspace{5mm}
% \subfigure[Spectrum at Earth]
%   {
\begin{figure}[!t]
	\centering
   \includegraphics[width=11cm]{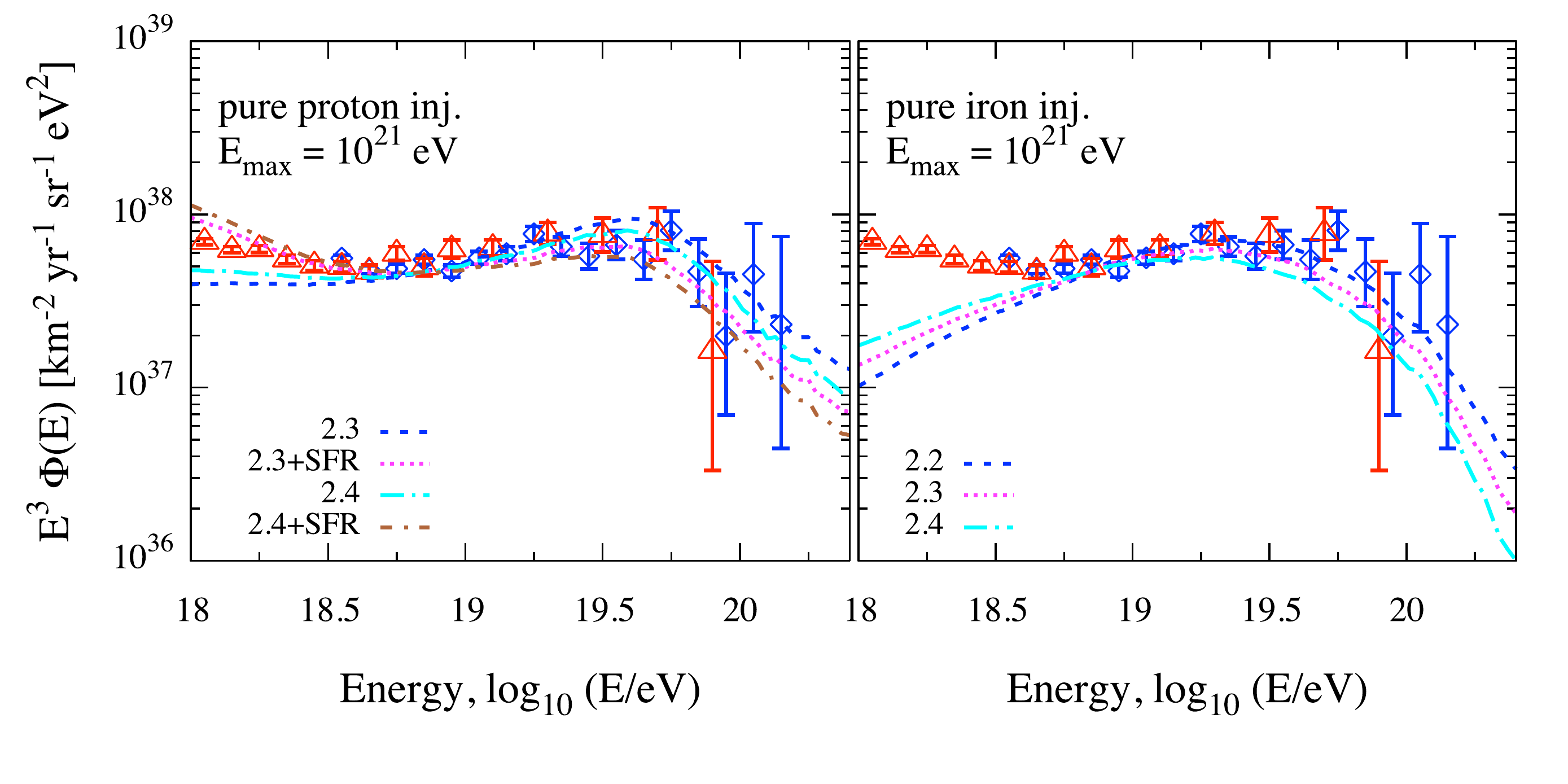}
  \caption{Expected all-particles energy spectra obtained from HERMES for different astrophysical scenarios, compared to observations reported by HiRes Collaboration (see the text). The legend indicates the spectral index at the source and the source evolution adopted (only star formation rate, in this case).}
   \label{fig:HERMES-spectra}
\end{figure}
%   }	
%   \caption{\textbf{(a)} Surviving probability of protons (see the text) produced by an homogenous distribution of sources within 300~Mpc, as a function of the propagation distance and for different energy threshold $E_{\text{thr}}$ at Earth. A power-law injection with energy cutoff $10^{21}$~eV and spectral index $2.4$ is used. We show the result of our simulations performed with HERMES (solid line), compared to those obtained with CRPropa~v1.4 (dashed line) and by Allard (dotted line). \textbf{(b)} Expected all-particles energy spectra obtained from HERMES for different astrophysical scenarios, compared to observations reported by HiRes Collaborations (see the text).}
%\end{figure*}

Successively, we estimate the GZK horizon for both protons and iron nuclei, as a function of the energy threshold at Earth. In particular, we compare against well-known results in literature~\cite{harari2006ultrahigh} and CRPropa~v2.0$\beta$\footnote{The version used here is dated September 2011.}, the up-to-date version of the Monte Carlo code simulating the 3D propagation of nuclei in a magnetized Universe~\cite{sigl2011icrc,kampert2013crpropa}. In Fig.\,\ref{fig-hermes-comp4} we show the GZK horizon of protons (left panel) and iron nuclei (right panel). In both cases, the horizons obtained by HERMES are in agreement with those of CRPropa over the whole energy range under consideration, although for iron nuclei some differences are present at the lowest energy. 

Moreover, we estimate the expected energy spectra of UHECR at Earth in different astrophysical scenarios, involving evolution of sources, different spectral indices and mass composition at the source. The result, shown in Fig.\,\ref{fig:HERMES-spectra}, are compared against recent observations reported by the HiRes Collaboration \cite{hires2008spectrum}. For sake of simplicity, we show only some representative spectra: a study of their goodness in reproducing the observed UHECR spectrum is beyond the scope of the present paper and it will be the subject of a future study.

\begin{figure}[!t]
\centering
\subfigure[2MRS Catalogue]
   {
	  \includegraphics[width=6.6cm]{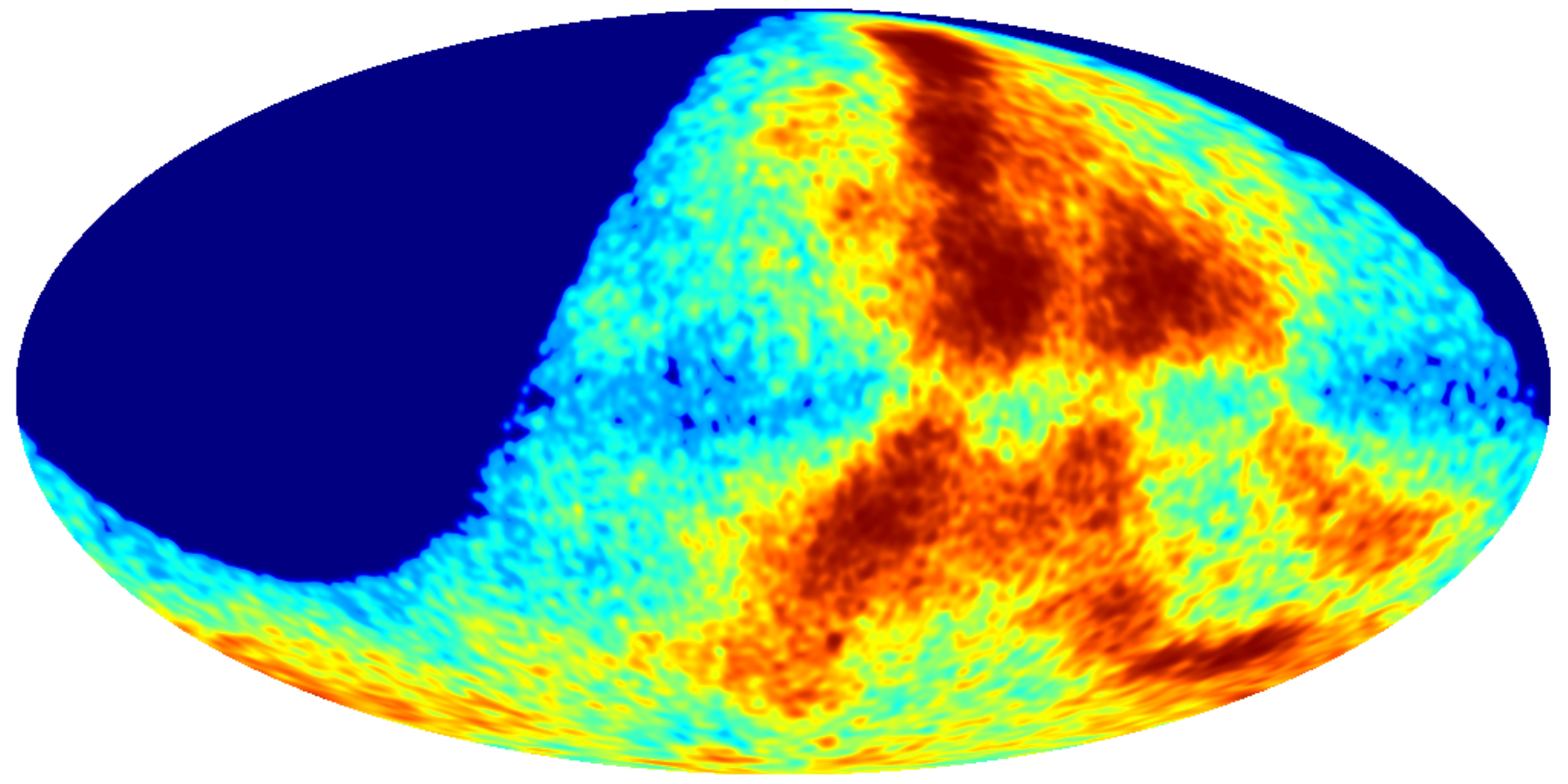}
	  \label{fig:skymap2MRS}
   }
   \hspace{5mm}
\subfigure[SWIFT58 Catalogue]
   {
	  \includegraphics[width=6.6cm]{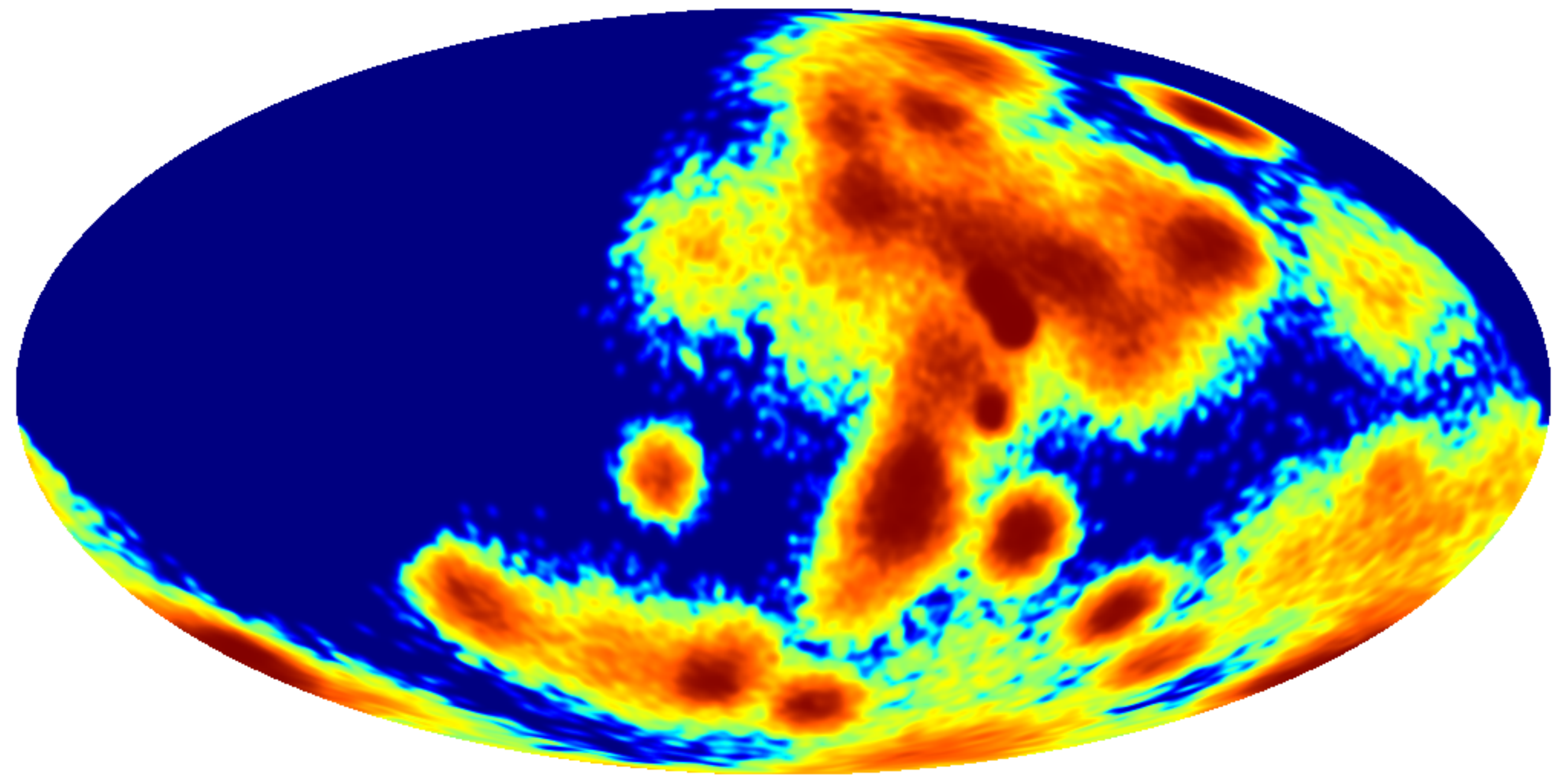}
	  \label{fig:skymapSWIFT}
   }
\subfigure[2MRS Catalogue + Isotropic]
   {
	  \includegraphics[width=6.6cm]{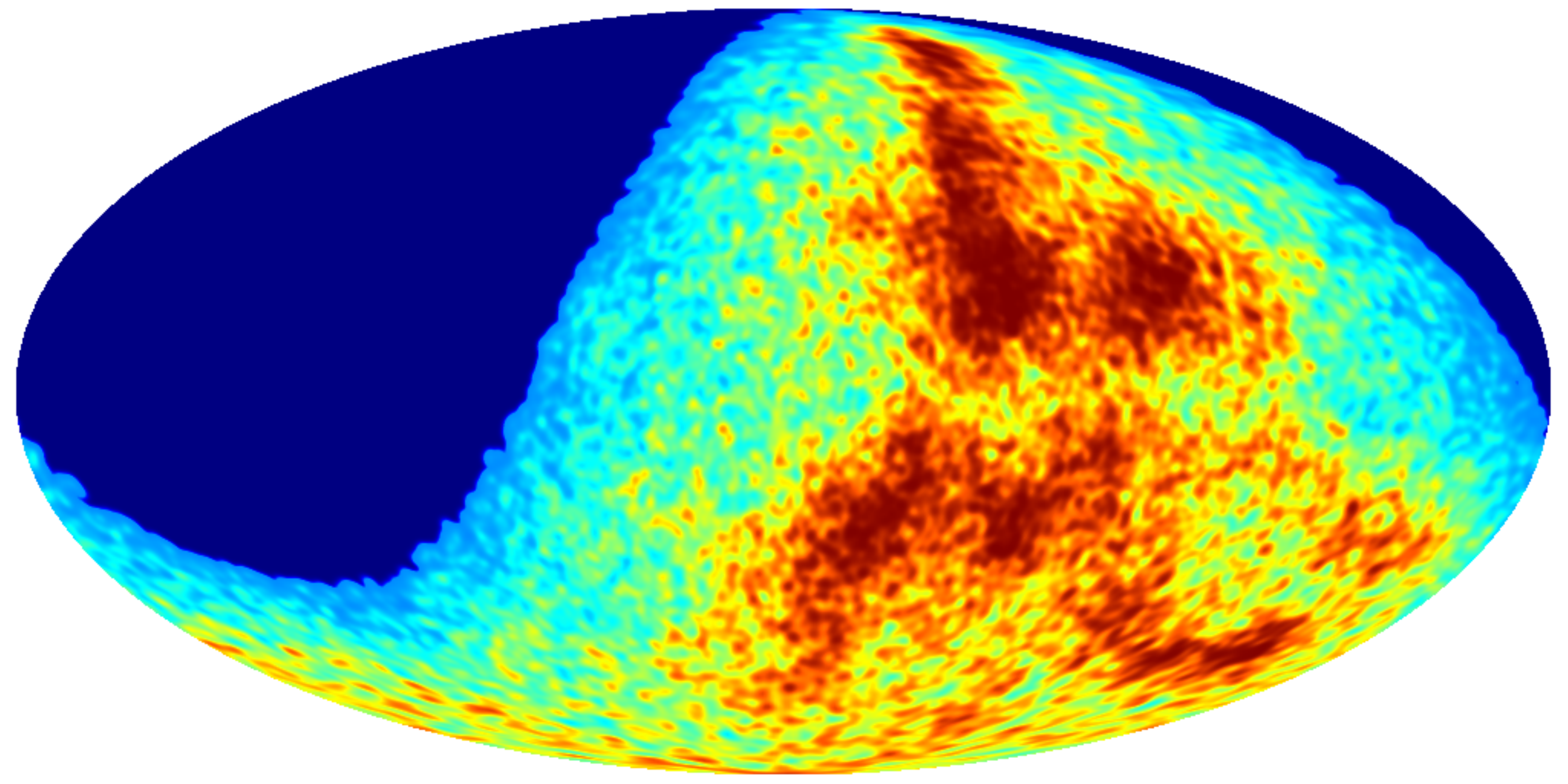}
	  \label{fig:skymap2MRSiso}
   }
   \hspace{5mm}
\subfigure[SWIFT58 Catalogue + Isotropic]
   {
	  \includegraphics[width=6.6cm]{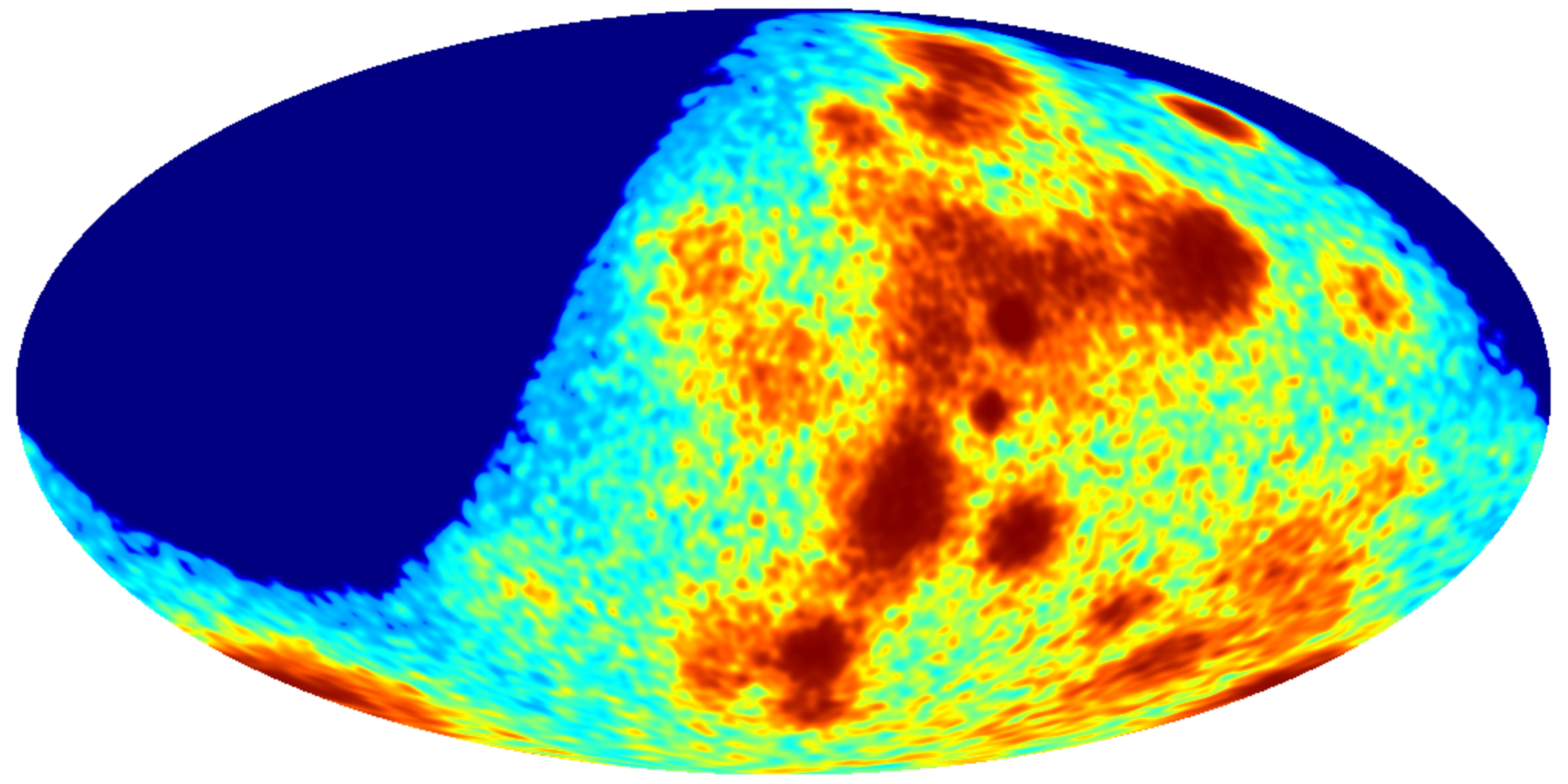}
	  \label{fig:skymapSWIFTiso}
   }

\caption{Skymaps, accounting for the Pierre Auger Observatory non-uniform exposure, of simulated UHE protons produced by nearby sources (within 200~Mpc) experiencing deflections due to an intervening extragalactic magnetic field. Galactic coordinates are shown. 2MRS (\ref{fig:skymap2MRS} and \ref{fig:skymap2MRSiso}) and SWIFT-BAT 58-months (\ref{fig:skymapSWIFT} and \ref{fig:skymapSWIFTiso}) are considered. See the text for further details.}
\end{figure}

As a final application, we simulated protons from real candidate sources in the nearby Universe, with distance between 4 and 200~Mpc. We included the effect of deflections due to an intervening Kolmogorov-like extragalactic magnetic field with r.m.s. strength of 2~nG and coherence length of 1~Mpc. Moreover, we consider the case of absence of isotropic contamination and the case where simulation are contaminated with 56\% isotropic events, according to recent measurements of the Pierre Auger Collaboration \cite{auger2010correlation}. The resulting skymaps of simulated events, as they would be observed by accounting for the non-uniform exposure of the Pierre Auger Observatory, are shown in Fig.\,\ref{fig:skymap2MRS} and Fig.\,\ref{fig:skymap2MRSiso}, for candidate sources of UHECR from 2MASS Redshift Survey \cite{huchra20112mass} with magnitude ranging from -27.5 to -9.8, and in Fig.\,\ref{fig:skymapSWIFT} and Fig.\,\ref{fig:skymapSWIFTiso} for active galactic nuclei from SWIFT-BAT 58months \cite{swift2010}.

Although a deeper analysis of correlation and intrinsic clustering is out of the scope of this paper, the results show how HERMES can be used for such purposes. Moreover, it is possible to investigate the compatibility between simulated scenarios and observation by coupling HERMES with other methods. For instance, it is possible to quantify the clustering signal in the arrival direction distribution \cite{dedomenico2011multiscale} or to perform multi-messenger analysis including photons propagated with EleCa \cite{settimo2012eleca}. Another interesting application is to use the parameterization based on the generalized Gumbel distribution \cite{dedomenico2013reinterpreting} to perform detailed mass composition studies, as comparing the expected first and second momenta of the $\text{X}_{\text{max}}$ distribution from different scenarios against observations.

%%%%%%%%%%%%%%%%%%%%%%%%%%%%%%%%%%%%%%%%%%%%%%%%%%%%%%
\section{Conclusions and outlook}

Realistic simulations of the propagation of UHECR might help to shed light on their origin and their nature. In this work, we presented HERMES, the \emph{ad hoc} Monte Carlo code we have developed to propagate UHECR in a magnetized Universe. We have briefly discussed the theoretical framework behind HERMES, involving the modeling of cosmology, magnetic fields, nuclear interactions between UHECR and relic photons of the extragalactic background radiation, and the production of secondary particles. The distribution of sources, their intrinsic luminosity, injection spectrum and evolution are tunable parameters in HERMES, allowing to simulate a wide variety of astrophysical scenarios and to investigate the impact of propagation on physical observable as the flux, or the chemical composition observed at Earth. 

We showed some representative applications validating the suitability of HERMES for astroparticle studies at the highest energies. More specifically, we estimated the surviving probability of UHE protons, the GZK horizons of nuclei, the all-particle spectrum observed at Earth in different astrophysical scenarios and the expected arrival direction distribution of UHECR produced from different catalogues of nearby candidate sources.

The major advantage in using HERMES is in its modularity, allowing high customization of involved physical and astrophysical parameters. In fact, it is possible, for instance, to add new models of extragalactic background radiations or nuclear interactions, according to up-to-date measurements.

In the near future, we will release a stable version of our simulator for public use and, in the meanwhile, we will make available for the community libraries of propagated nuclei useful for mass composition and energy spectrum analysis.

%Improved observations, together with modeling of both sources and propagation coupled to clustering correlation analysis, might be able to open new insights to the solution of the still unsolved UHECR puzzle.

\begin{acknowledgement}
The author is in debt with the Pierre Auger Collaboration for the invaluable discussions and acknowledges the financial support of the Scuola Superiore di Catania, the Department of Physics and Astronomy of the University of Catania and INFN (Sez.\,Catania). The author would like to thank P.L. Ghia for invaluable support and precious suggestions, and H. Lyberis and M. Settimo for useful discussions before, during and after the realization of HERMES.
\end{acknowledgement}

\bibliographystyle{jcappub}
%\bibliography{epjp_bruno_rossi}

\providecommand{\href}[2]{#2}\begingroup\raggedright\endgroup

\end{document}